\documentclass[twocolumn,english,aps,superscriptaddress,showpacs,prl,floatfix,longbibliography]{revtex4-1}
\usepackage{color}
\usepackage{babel}
\usepackage{amsmath}
\usepackage{amssymb}
\usepackage{graphicx}
\usepackage{esint}
\usepackage[unicode=true,pdfusetitle,bookmarks=true,bookmarksnumbered=false,bookmarksopen=false,breaklinks=true,pdfborder={0 0 0},backref=false,colorlinks=true]{hyperref}
\hypersetup{linkcolor=blue,citecolor=blue,urlcolor=blue}
\usepackage{breakurl}

\makeatletter
\usepackage{color}

\makeatother

\newcommand{\lam}{\alpha}    
\newcommand{\edif}{\lambda}   
\newcommand{\kap}{\kappa}     
\newcommand{\xc}{x_{\rm cr}}  
\newcommand{\hc}{h_{\rm cr}}  

\begin{document}
\title{Cluster-glass phase in pyrochlore XY antiferromagnets with quenched disorder}

\author{Eric C. Andrade}
\affiliation{Instituto de F\'{i}sica de S\~ao Carlos, Universidade de S\~ao Paulo, C.P. 369,
S\~ao Carlos, SP,  13560-970, Brazil}

\author{Jos\'e A. Hoyos}
\affiliation{Instituto de F\'{i}sica de S\~ao Carlos, Universidade de S\~ao Paulo, C.P. 369,
S\~ao Carlos, SP,  13560-970, Brazil}

\author{Stephan Rachel}
\affiliation{Institut f\"ur Theoretische Physik, Technische Universit\"at Dresden,
01062 Dresden, Germany}
\affiliation{School of Physics, University of Melbourne, Parkville, VIC 3010, Australia}

\author{Matthias Vojta}
\affiliation{Institut f\"ur Theoretische Physik, Technische Universit\"at Dresden,
01062 Dresden, Germany}

\begin{abstract}
We study the impact of quenched disorder (random exchange couplings
or site dilution) on easy-plane pyrochlore antiferromagnets. In the
clean system, order-by-disorder selects a magnetically ordered state
from a classically degenerate manifold. In the presence of randomness,
however, different orders can be chosen locally depending on details
of the disorder configuration. Using a combination of analytical considerations
and classical Monte-Carlo simulations, we argue that any long-range-ordered
magnetic state is destroyed beyond a critical level of randomness
where the system breaks into magnetic domains due to random exchange
anisotropies, becoming, therefore, a glass of spin clusters, in accordance
with the available experimental data. These random anisotropies originate from off-diagonal exchange couplings in the microscopic Hamiltonian, establishing their relevance to other magnets with strong spin-orbit coupling.

\end{abstract}

\date{\today}
\maketitle


Rare-earth pyrochlores form one of the most interesting families of
frustrated magnets. A lattice of corner-sharing tetrahedra combined
with a multiplicity of crystal-field effects for rare-earth ions~\cite{gardner10}
gives rise to a plethora of novel states~\cite{yan17}. Among them
are disordered spin ices~\cite{harris97,moessner98,henley10,castelnovo11}
and quantum spin liquids~\cite{hermele04,ross11,gingras14}, found
in materials with magnetic easy-axis anisotropy. In contrast, compounds
exhibiting an easy-plane (or XY) anisotropy tend to order antiferromagnetically~\cite{champion03,zhitomirsky12,savary12,zhitomirsky14,mcClarty14,javanparast15,hallas17}.
A number of them realize an ``order-by-disorder'' mechanism where
a long-range-ordered state is selected, via either thermal or quantum
fluctuations, from a classically degenerate manifold resulting from
strong frustration~\cite{villain80,shender82,henley89,savary12}.
In the parameter regime relevant to the paradigmatic example Er$_{2}$Ti$_{2}$O$_{7}$,
both classical and quantum fluctuations select the noncoplanar state
dubbed $\psi_{2}$ from a one-parameter manifold, in a remarkable
agreement with experiments~\cite{zhitomirsky12,savary12}.

Quenched disorder provides a different route for lifting
the classical degeneracy by locally relieving the frustration~\cite{villain79}.
Previous theoretical studies showed that both bond randomness
and site dilution tend to stabilize, for small disorder, the coplanar
state dubbed $\psi_{3}$~\cite{maryasin14,andreanov15}. This insight
motivated a series of experiments in inhomogeneous XY pyrochlore magnets.
In Er$_{2-x}$Y$_{x}$Ti$_{2}$O$_{7}$~\cite{gaudet16} magnetic
$\mbox{Er}^{3+}$ is substituted by nonmagnetic $\mbox{Y}^{3+}$,
corresponding to site dilution. For NaCaCo$_{2}$F$_{7}$~\cite{ross16,sarkar17}
and NaSrCo$_{2}$F$_{7}$~\cite{ross17} quenched disorder arises
from site mixing on the pyrochlore A sites (Na/Ca and Na/Sr, respectively),
with the leading effect on magnetism being bond randomness. However,
all experiments find either the $\psi_{2}$ state, for small disorder,
or short-range magnetic correlations below a freezing temperature
$T_{f}$ at stronger disorder, suggesting a spin-glass state.

\begin{figure}[tb]
\begin{centering}
\includegraphics[clip,width=0.45\columnwidth]{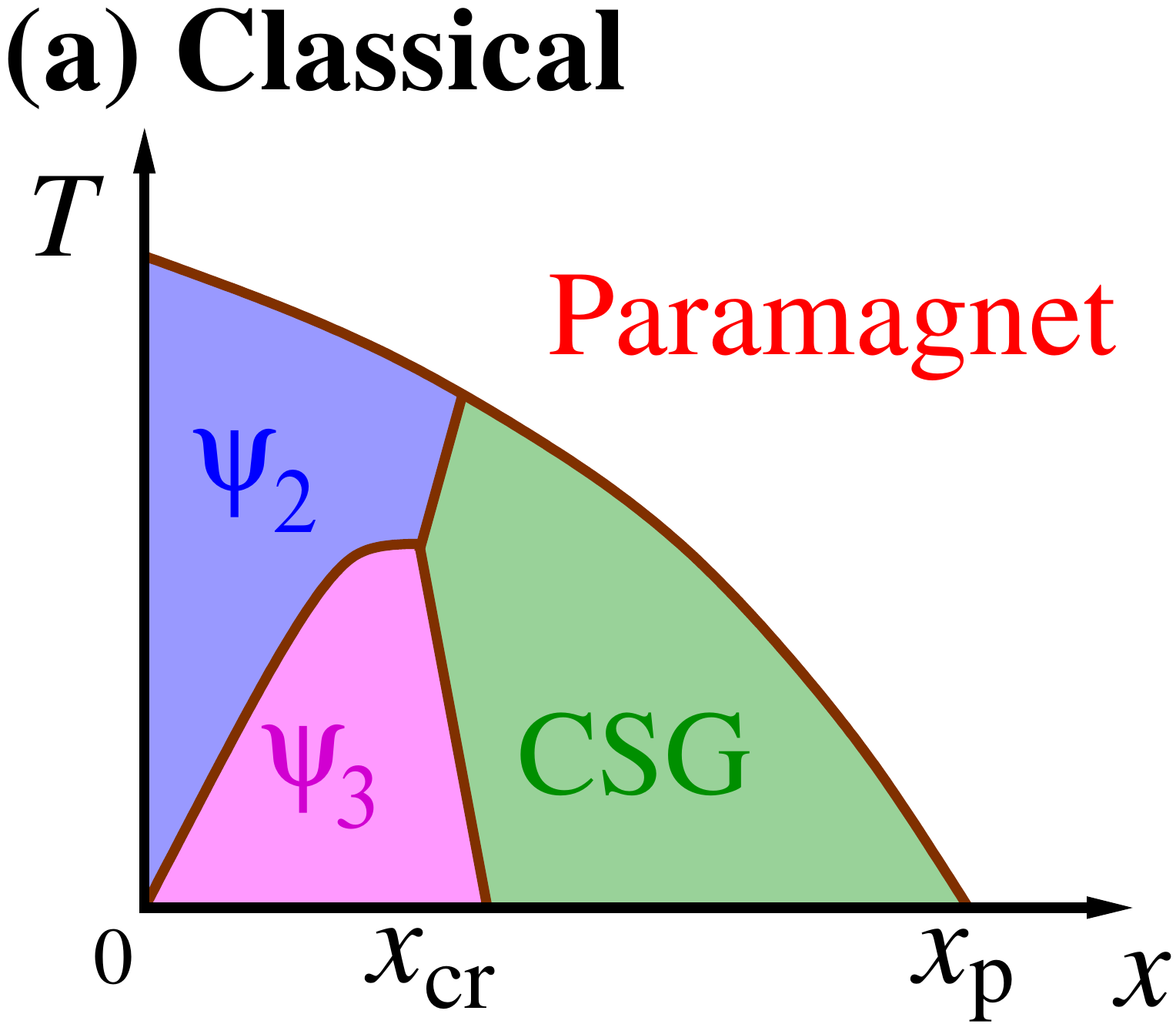}\hfill\includegraphics[clip,width=0.45\columnwidth]{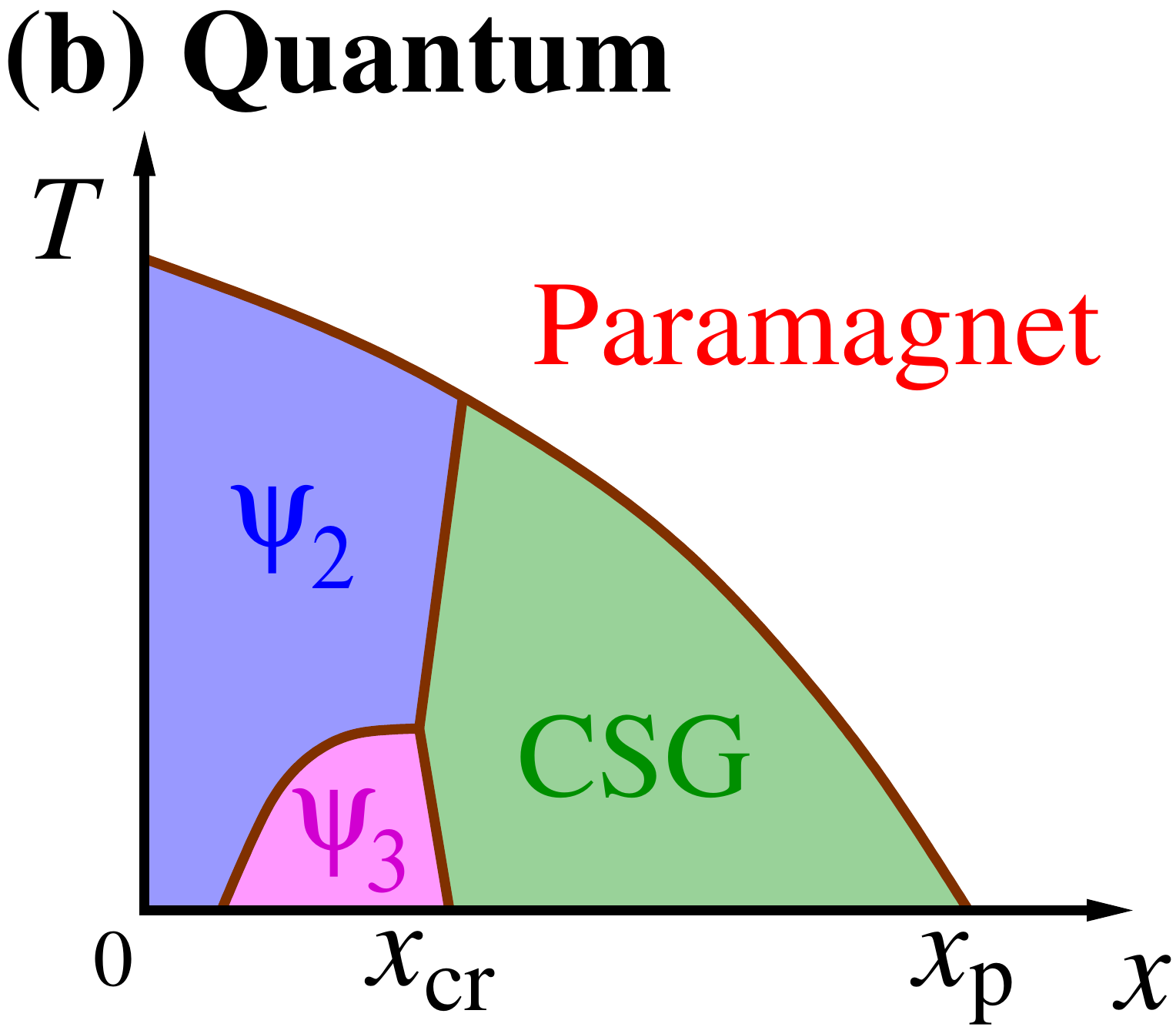}
\par\end{centering}
\caption{(Color online) Schematic phase diagram for the easy-plane pyrochlore
antiferromagnet \eqref{eq:Hxy} as function of temperature $T$ and
site dilution $x$, for parameters with $0<J_{\pm\pm}/J_{\pm}\le2$
where order-by-disorder selects the $\psi_{2}$ state in the clean
limit, $x\!=\!0$. Quenched disorder tends to select the $\psi_{3}$
state at low $T$, but both $\psi_{2,3}$ are destroyed beyond a critical
level of disorder $\xc$ where randomness breaks the system into magnetic
domains resulting in a cluster spin glass (CSG) phase which survives
up to the percolation threshold $x_{{\rm p}}$. (a) Without quantum
fluctuations, i.e., in the classical limit. (b) With quantum fluctuations.
Here $\psi_{3}$ may disappear completely, for details see text.
}
\vspace*{-10pt}
\label{fig:PD}
\end{figure}

In this Letter, we solve this puzzle. We develop a more general theory, valid in the experimentally relevant regime, showing that disorder-induced random anisotropies destabilize magnetic long-range order (LRO) of easy-plane pyrochlores. We derive a semi-quantitative stability criterion which indicates that LRO is destroyed beyond a critical level of quenched disorder.
Using extensive Monte-Carlo (MC) simulations for the relevant classical
model, we verify the tendency towards $\psi_{3}$ order in the weakly
diluted regime \cite{maryasin14,andreanov15}, and provide compelling
evidence that the low-$T$ state at stronger disorder is a cluster
spin-glass (CSG) phase where the system breaks into domains (spins
clusters) exhibiting a variety of {\em local} ordering patterns
besides the $\psi_{2}$ and $\psi_{3}$ ones. For the quantum case,
we argue that the tendency towards $\psi_{3}$ order is diminished,
such that a significant portion of the phase diagram is dominated
by the CSG phase, Fig.~\ref{fig:PD}, in agreement with available
experiments. 


\emph{Model}.--- The low-temperature properties of many rare-earth
insulating pyrochlore oxides are well described by an effective spin-$1/2$
model with anisotropic exchange interactions due to the combination
of spin-orbit and crystalline electric-field effects~\cite{gardner10}.
In Er$_{2}$Ti$_{2}$O$_{7}$, for instance, the spins have a dominant
planar nature, and the associated nearest-neighbor anisotropic XY
model~\cite{curnoe08,thompson11,ross11,zhitomirsky12,savary12} can
be written as
\begin{equation}
\mathcal{H}=-\sum_{\left\langle jk\right\rangle }\left[J_{jk}^{xx}S_{j}^{x}S_{k}^{x}+J_{jk}^{yy}S_{j}^{y}S_{k}^{y}+J_{jk}^{xy}\left(S_{j}^{x}S_{k}^{y}+S_{j}^{y}S_{k}^{x}\right)\right].\label{eq:Hxy}
\end{equation}
Here, the sum runs over pairs of nearest-neighbor (NN) sites on a
cubic pyrochlore lattice, the couplings are
\begin{equation}
J_{jk}^{xx(yy)}=2(J_{jk}^{\pm}\mp J_{jk}^{\pm\pm}\mbox{cos}\theta_{jk}),\ J_{jk}^{xy}=2J_{jk}^{\pm\pm}\mbox{sin}\theta_{jk},\label{eq:exchange}
\end{equation}
and the spin components $S^{x,y}$ are written in the local coordinate
reference frames (one for each of the four sublattices) which are
confined to planes perpendicular to the local $\left\langle 111\right\rangle $
axes. $J^{\pm}$ and $J^{\pm\pm}$ are the symmetry-allowed NN exchange
couplings~\cite{ross11,savary12,savary12a}. We adopt the choice
of the local frames as in Refs.~\onlinecite{sarkar17,maryasin14},
with the corresponding angular phases (inside a tetrahedron of sites
labeled from $0$ to $3$) $\theta_{01}=\theta_{23}=0$, $\theta_{02}=\theta_{13}=2\pi/3$,
and $\theta_{03}=\theta_{12}=-2\pi/3$.

In the clean limit, and for $-2<\lam\equiv J^{\pm\pm}/J^{\pm}<2$,
the mean-field ground states of \eqref{eq:Hxy} are given by spins
collectively pointing along any direction in the XY plane, exhibiting
a continuous U(1) degeneracy with energy $E_{0}/J^{\pm}=-6NS^{2}$.
The order-by-disorder mechanism selects a finite set of six states
out of the degenerate manifold: for $0<\lam<2$ ($-2<\lam<0$) the
$\psi_{2}$ ($\psi_{3}$) state is selected which corresponds to spins
pointing along one of the $\cos\left(\frac{\pi}{3}n\right)\hat{x}+\sin\left(\frac{\pi}{3}n\right)\hat{y}$
($\cos\left(\frac{\pi}{3}n+\frac{\pi}{6}\right)\hat{x}+\sin\left(\frac{\pi}{3}n+\frac{\pi}{6}\right)\hat{y}$)
directions, with $n=0,\dots,5$, in the local reference frame~\cite{zhitomirsky12,savary12}.

For definiteness, we introduce quenched disorder via
\begin{equation}
J_{jk}^{\pm}=J^{\pm}\left(1+\epsilon_{jk}\right),\:J_{jk}^{\pm\pm}=J^{\pm\pm}\left(1+\epsilon_{jk}\right),\label{eq:random-exchange}
\end{equation}
where $\epsilon_{jk}$ are random variables. Bond disorder corresponds
to $\epsilon_{jk}$ drawn independently from some distribution, while
site dilution yields $\epsilon_{jk}=-1$ if either site $j$ or site
$k$ hosts a vacancy and $\epsilon_{jk}=0$ otherwise. Site dilution
is parameterized by the concentration $x$ of non-magnetic impurities;
for bond randomness see~\cite{suppl}.


\emph{Destruction of order by random-fields effects.---} We adopt
a transparent argument by Aharony, originally constructed to show
the instability of an ordered magnetic state against weak random anisotropy~\cite{aharony78}.

We \emph{assume} LRO which is uniform in the local frames of the Hamiltonian
\eqref{eq:Hxy}, with $\left|\lam\right|<2$, such that $\left\langle \mathbf{S}\right\rangle =\left\langle S^{x}\right\rangle \hat{x}+\left\langle S^{y}\right\rangle \hat{y}=m\hat{n}_{\parallel}$
where $\left\langle \cdots\right\rangle $ denotes the thermal average.
The corresponding local exchange field is $\mathbf{h}_{j}=\sum_{k=1}^{z}(J_{jk}^{xx}\left\langle S^{x}\right\rangle +J_{jk}^{xy}\left\langle S^{y}\right\rangle )\hat{x}+(J_{jk}^{yy}\left\langle S^{y}\right\rangle +J_{jk}^{xy}\left\langle S^{x}\right\rangle )\hat{y}$,
with the sum running over all the $z=6$ NN sites. In the presence
of random off-diagonal disorder the local exchange field $\mathbf{h}_{j}=h_{j}^{\parallel}\hat{n}_{\parallel}+h_{j}^{\perp}\hat{n}_{\perp}$
is not parallel to the mean magnetization. If we
assume, for instance, $\left\langle S^{x}\right\rangle =\frac{1}{2}$
(and $\left\langle S^{y}\right\rangle =0$) the local transverse
component stems from the coupling $J_{jk}^{xy}$ in \eqref{eq:Hxy},
and is simply given by $h_j^{\perp}/J^{\pm\pm}=\sum_{k=1}^{6}\epsilon_{jk}\sin\theta_{jk}$,
see also Eqs. \eqref{eq:exchange} and \eqref{eq:random-exchange}.

The random transverse field $h_{j}^{\perp}$ \cite{fieldvsaniso} tips the local magnetization
away from the mean direction $\hat{n}_{\parallel}$. The resulting
transverse magnetization can be estimated in linear response as $\left\langle S_{j}^{\perp}\right\rangle =\sum_{k}\chi_{jk}^{\perp}h_{k}^{\perp}$
where $\chi_{jk}^{\perp}$ is the transverse bulk susceptibility of
the clean system. The disorder-averaged transverse magnetization,
$\overline{\langle S_{j}^{\perp}\rangle}$, vanishes because $h_{j}^{\perp}$
has zero mean. In contrast, the averaged magnetization correlation
function is non-zero: $\overline{\left\langle S_{i}^{\perp}S_{j}^{\perp}\right\rangle }\propto(\delta h)^{2}\int{\rm d}^{d}q\left(\chi^{\perp}(\mathbf{q})\right)^{2}e^{i\mathbf{q}\cdot\mathbf{r}_{ij}}$
where $\left(\delta h\right)^{2}\equiv\overline{h_{i}^{\perp2}}>0$
~\footnote{We have assumed uncorrelated disorder which generically applies to
the long-wavelength limit.} and $d=3$ the number of space dimensions. Further, $\chi^{\perp}(\mathbf{q})\sim1/(\edif+\kap_{\mu}q_{\mu}^{2})$
is the Fourier-transformed bulk susceptibility, with $\edif$ being
an effective anisotropy energy, such that the gap $\Delta\propto\sqrt{\edif}$,
and $\kap_{\mu}$ parameterizing the gradient expansion \cite{savary12,suppl}.

Importantly, if the anisotropy energy $\edif$ \emph{were} zero, we
would have $\chi^{\perp}(\mathbf{q})\propto q^{-2}$, such that the
local transverse magnetization fluctuations diverge for $d\leq4$:
These fluctuations, arising from random off-diagonal exchange interactions
and transmitted by long-wavelength modes, then destroy the assumed
ordered state. This destruction of LRO can also be interpreted in
terms of breaking the system into domains of linear size $\ell$,
following Imry and Ma~\cite{imry_ma}. Consider domains inside which
the transverse exchange fields $\{h_{j}^{\perp}\}$ are atypically
strong, such that the local order parameter will align with it, gaining
an energy scaling as $\delta h\ell^{d/2}$ (as dictates the central
limit theorem). In addition, there is a domain-wall energy cost. $\edif\to0$
implies an (accidental) continuous symmetry, hence the order parameter
can be continuously distorted from the $\hat{n}_{\parallel}$ to the
$\hat{n}_{\perp}$ direction (in a region of fractions of $\ell$),
yielding a domain wall energy which scales as $J^{\pm}\ell^{d-2}$.
Thus, for $d<4$ it is favorable to break the system into domains
of linear size $\ell\gtrsim\left(J^{\pm}/\delta h\right)^{2/(4-d)}$.

If, instead, the anisotropy energy $\edif$ is finite, the divergence
is cured. Then the transverse spin fluctuations $\overline{\left\langle S_{i}^{\perp2}\right\rangle }$
remain small for small $\delta h$, i.e., the assumed LRO is stable
against weak randomness. A stability criterion can be obtained by
the condition $\overline{\left\langle S_{i}^{\perp2}\right\rangle }\ll1$.
This yields
\begin{equation}
\delta h\ll\kap^{d/4}\edif^{1-d/4}\label{crit}
\end{equation}
up to numerical prefactors \cite{suppl} where $\kap^{2}=\sum_{\mu}\kap_{\mu}^{2}/3$.
The criterion \eqref{crit} is consistent with the fact that small
randomness destroys LRO in the limit $\edif\to0$ for $d<4$, but
not for $d>4$. $\delta h$ grows with increasing randomness, and
hence we expect, for $\edif>0$ and $d=3$, the destruction of LRO
beyond a critical level of randomness.
Below we show that the resulting phase is a CSG.
Importantly, this argument is rather general, not restricted to the
specific choice of Eq. \eqref{eq:Hxy}, since it relies only on the existence of an off-diagonal exchange coupling \cite{aharony78}.


\emph{Effective anisotropy and critical disorder.---} Without quenched
disorder, the system selects the $\psi_{2}$ state (for $0<\lam<2$),
with the spin gap $\Delta$ generated by the order-by-disorder mechanism~\cite{zhitomirsky12,savary12,ross14}.
Hence, $\edif$ is generically finite, and the $\psi_{2}$ state is
stable against weak randomness, i.e., small $\delta h$. Indeed, a
perturbation generated by a single defect is expected to be screened
in the presence of a gap, such that a small concentration of defects
does not qualitatively change the bulk state \cite{maryasin14,andreanov15}.

Importantly, in the classical limit the $\psi_{2}$ gap arises exclusively
from thermal fluctuations, hence $\Delta\to0$ as $T\to0$. Quenched disorder
tends to stabilize the $\psi_{3}$ state instead \cite{maryasin14,andreanov15}, with the effective anisotropy energy $\edif$ scaling linearly with $x$
\cite{maryasin14}. Hence, the putative $\psi_{3}$ state has $\edif\propto x$
as $T\to0$. As we show in Ref.~\onlinecite{suppl}, the fluctuating
transverse field follows $\delta h\propto\sqrt{x}$, such that the
criterion \eqref{crit} is parametrically fulfilled for small $x$,
but can be expected to be violated at larger $x$. We conclude that,
in the classical limit and at low $T$, the $\psi_{3}$ state is stable
in a window $0<x<\xc$, but replaced by a CSG for $x>\xc$.
The quantum case is more involved and will be discussed later.

\begin{figure}
\begin{centering}
\includegraphics[width=0.5\columnwidth]{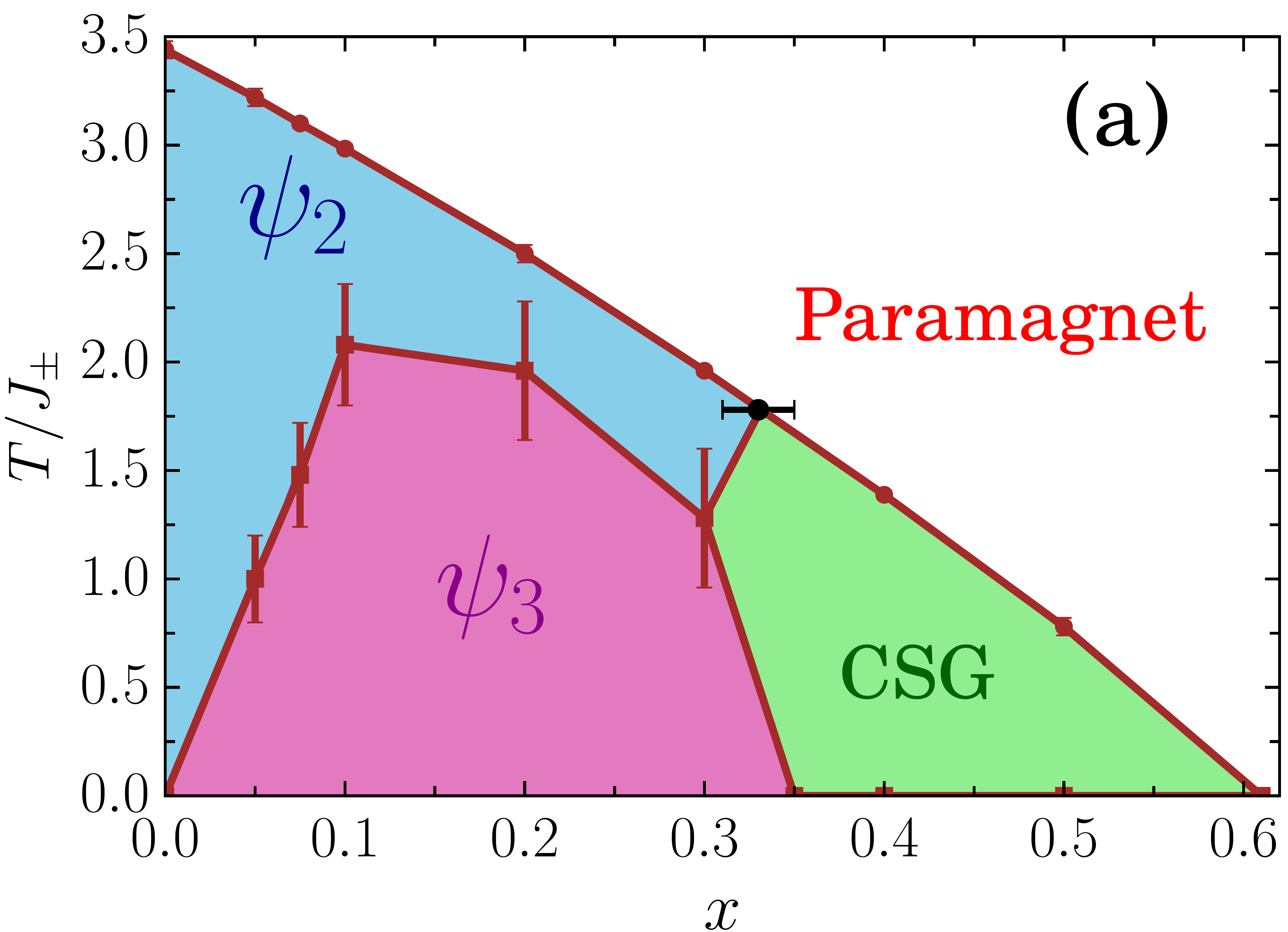}\includegraphics[width=0.5\columnwidth]{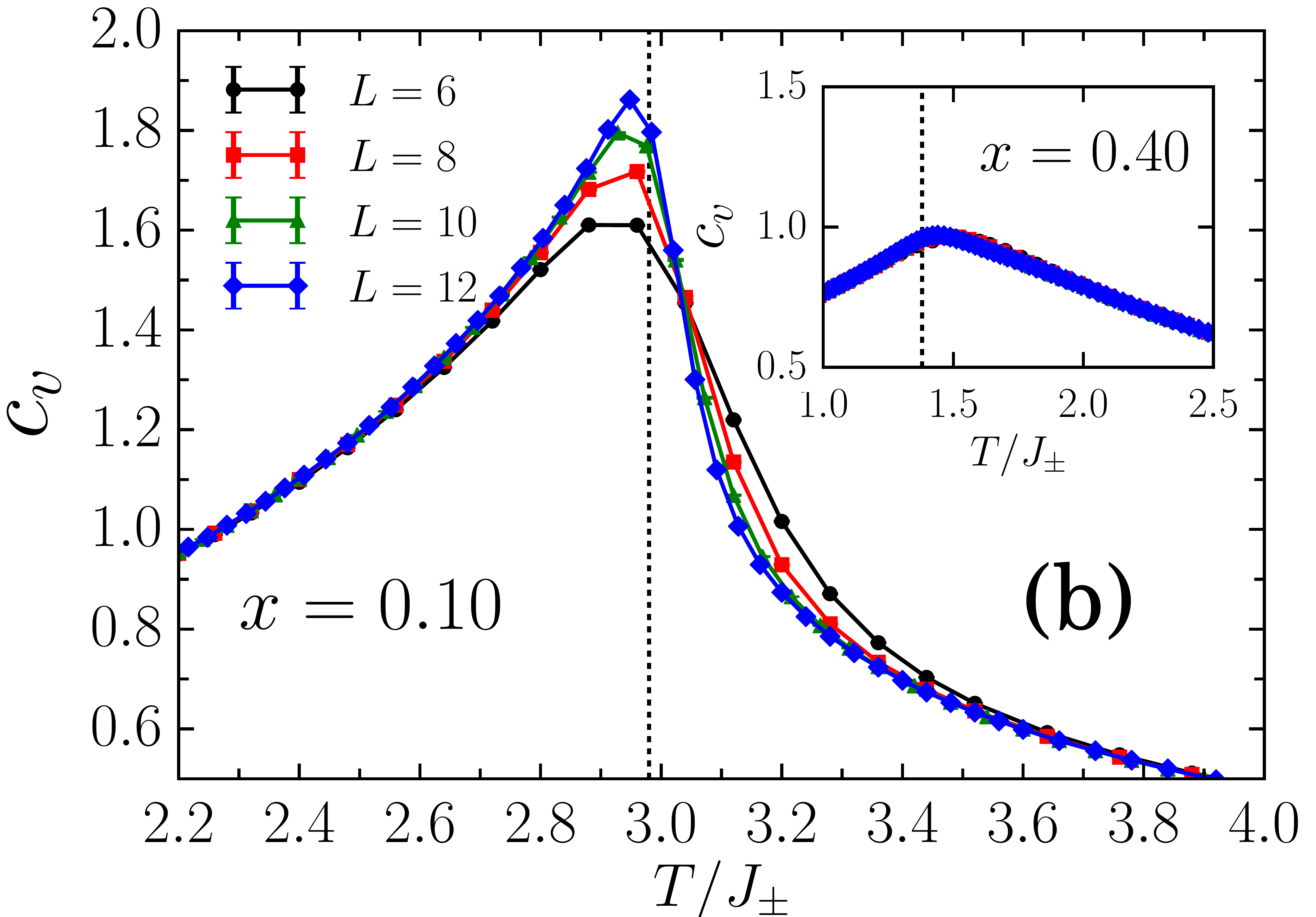}
\par\end{centering}

\begin{centering}
\includegraphics[width=0.5\columnwidth]{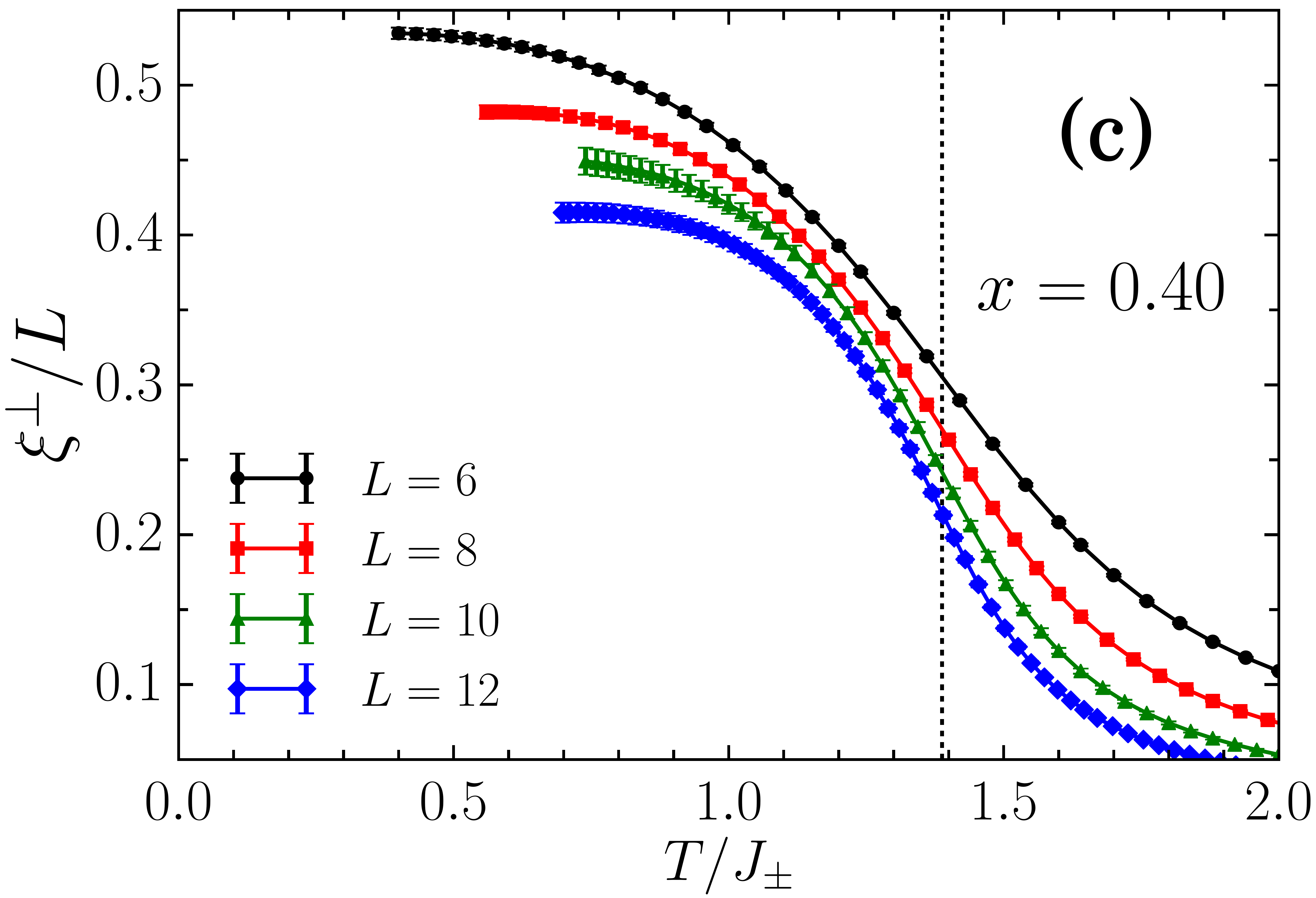}\includegraphics[width=0.5\columnwidth]{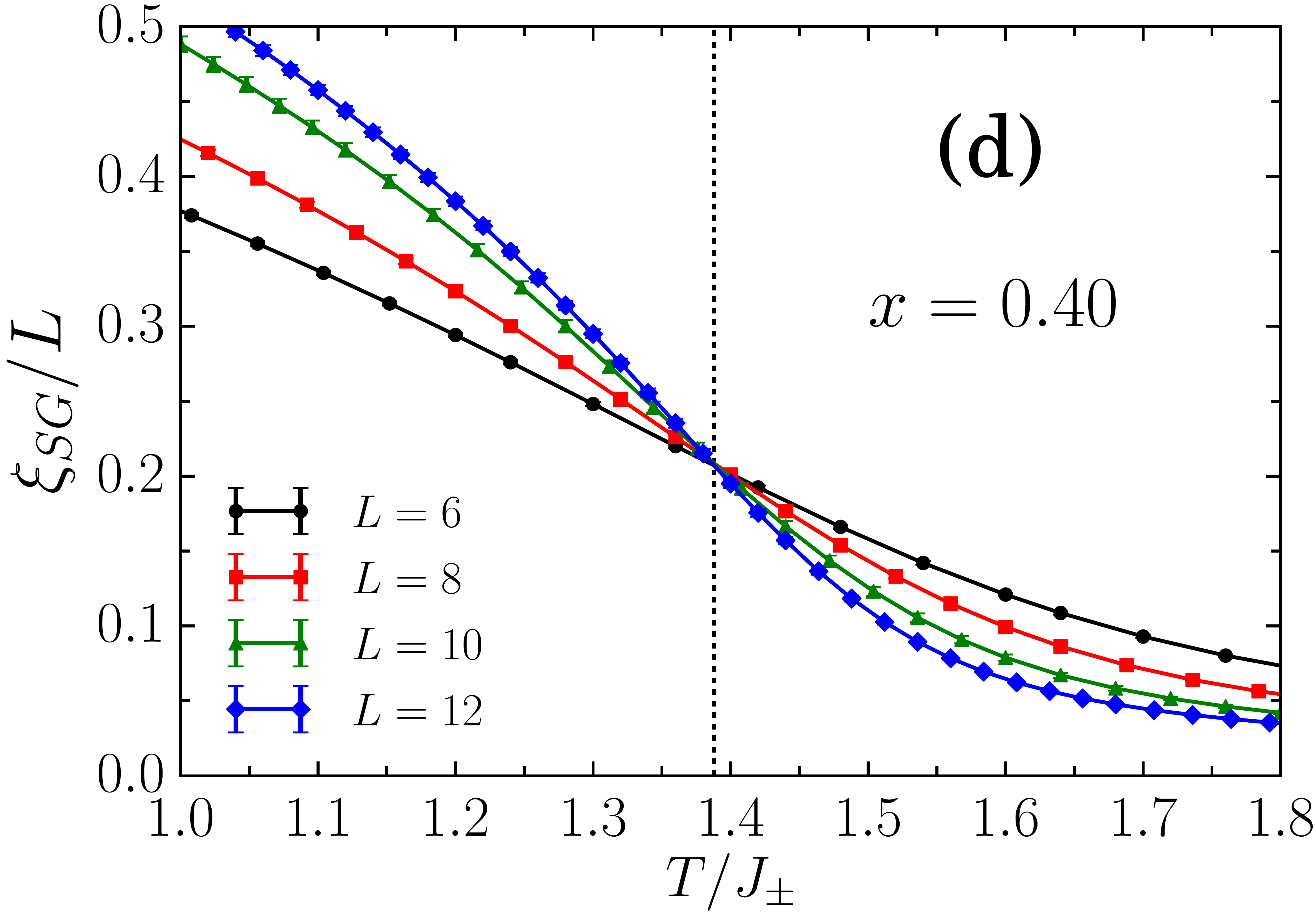}
\par\end{centering}

\begin{centering}
\includegraphics[width=0.5\columnwidth]{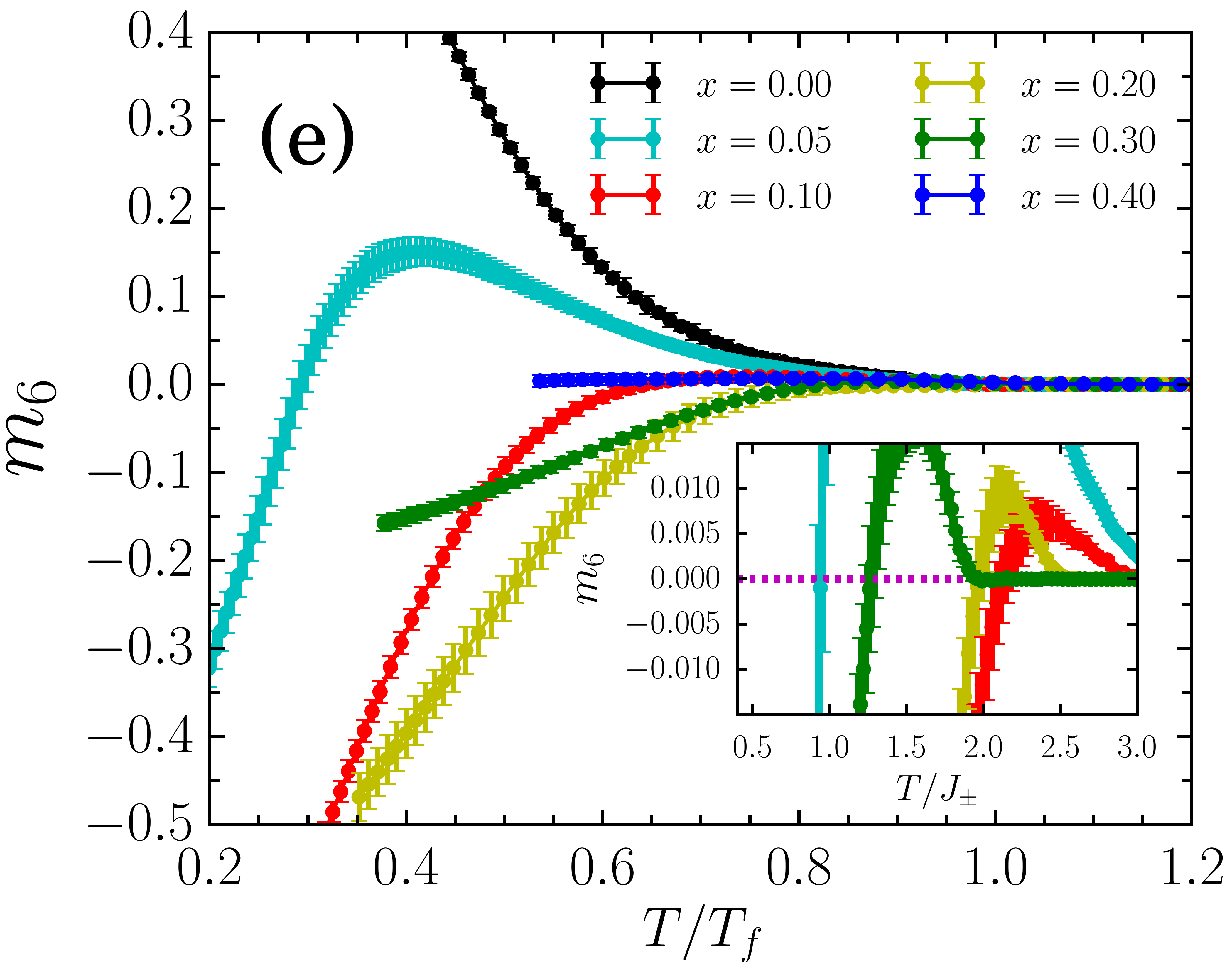}\includegraphics[width=0.5\columnwidth]{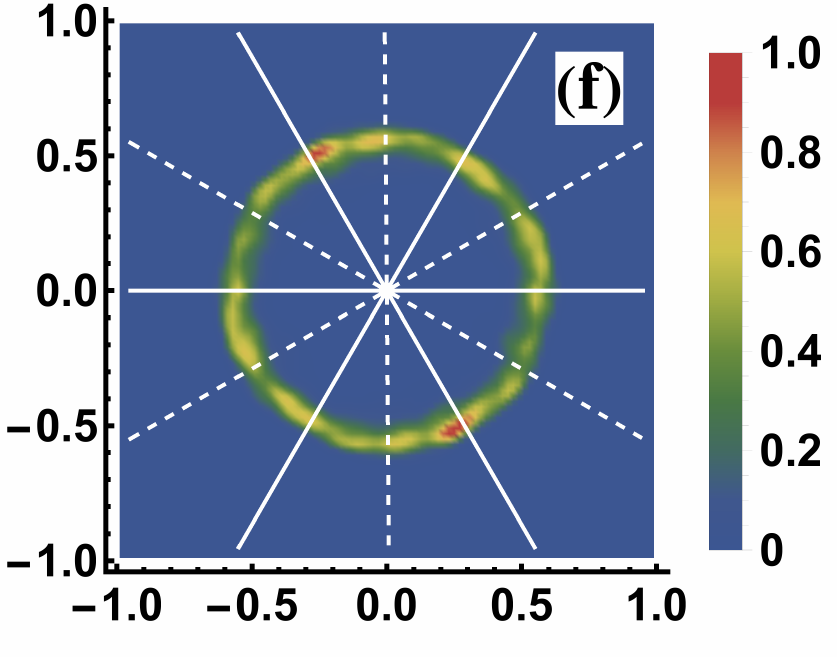}
\par\end{centering}

\caption{(Color online) Classical MC results for the Hamiltonian \eqref{eq:Hxy}
for off-diagonal exchange ratio $\lam=1$.
(a) Phase diagram as in Fig.~\ref{fig:PD}. The freezing temperature $T_{f}$ vanishes at the percolation threshold $x_{{\rm p}}=0.61$; the black dot marks the transition between $\psi_{2}$ and CSG, for details see text.
(b) Specific heat $c_{v}(T)$ for $x=0.1$ and $x=0.4$ (inset). The vertical dashed line marks $T_f$.
(c) Magnetic correlation length, plotted as $\xi^{\perp}(T)/L$, for $x=0.4$ showing no crossing points.
(d) Spin-glass correlation length, plotted as $\xi_{{\rm SG}}(T)/L$, for $x=0.4$ showing
a crossing point at $T_{f}/J_{\pm}=1.39(2)$.
(e) Clock order parameter $m_{6}$ as a function of $T/T_{f}$ for $L=10$ and several
values of $x$. Inset: zoom around the region where $m_{6}$ changes
sign.
(f) Sample-to-sample distribution function $P\left(m_{x},m_{y}\right)$
at $T=T_{f}/2$ for $x=0.4$ and $L=12$. The full (dashed) radial
lines show the expected positions of the peaks associated to the $\psi_{2(3)}$
states.}
\label{fig:fig2}
\end{figure}

\emph{Classical phase diagram.---}
We are now in the position to discuss the classical phase diagram of the model \eqref{eq:Hxy}. As originally explained in Refs.~\onlinecite{maryasin14,andreanov15}, the behavior at small $x$ is governed by the competition between $\psi_{2}$ and $\psi_{3}$ LRO, favored by thermal fluctuations and weak disorder, respectively. This results in a phase boundary varying linearly
with $x$. With increasing $x$, random-field effects grow, and LRO is
eventually destroyed in favor of a CSG at all $T$.
The low-$T$ competition between $\psi_{3}$ and CSG is based on energetics, such that their phase boundary depends weakly on $x$ at low $T$.
Further, the effective anisotropy $\edif$ is particularly small near the $\psi_{2}$--$\psi_{3}$ boundary, and this is where CSG will win first upon increasing $x$.
Together, these considerations yield the qualitative phase diagram in Fig.~\ref{fig:PD}(a),
and they are well borne out by our quantitative numerical simulations, Fig.~\ref{fig:fig2}.


\emph{Classical MC simulations.---} We turn to a detailed analysis
of the state at large disorder. Previous theoretical investigations
of related cases~\cite{fisher85,proctor14,proctor15} suggest a glassy
state: The system breaks into domains of size $\ell$, exhibiting
no long-range magnetism, and eventually freezes into a spin glass
below a temperature $T_{f}$. This scenario is in accordance with
available experimental results for the random XY pyrochlores~\cite{gaudet16,ross16,sarkar17,ross17},
and we now provide numerical evidence for it in the classical limit~\footnote{The present glass state is fundamentally different from those proposed
for the Ising and the isotropic Heisenberg pyrochlores \cite{saunders07,andreanov10}.
In those examples, the glassiness is due to effective moments which
interact via a spin-liquid background. In contrast, in our case the
CSG is due to domain formation.}.

We perform classical MC simulations of the model \eqref{eq:Hxy} in
the presence of site dilution and bond randomness~\cite{suppl}.
Interpreting the simulation results requires care due to the several
length scales present in the problem. In the clean limit, besides
the linear system size $L$, there is an emergent length $\Lambda(T)$~\cite{lou07,wenzel11}
associated to a dangerously irrelevant $\mathbb{Z}_{6}$ anisotropy~\cite{zhitomirsky14}.
Therefore, the ground-state selection only takes place for $L\gg\Lambda$;
to observe this numerically requires either low $T$ or large $L$.
With quenched disorder there is yet another length scale, the domain
size $\ell$, and we require $L\gg\ell$ to observe domain formation.

To monitor the order in the local XY planes we compute $m_{x\left(y\right)}=N^{-1}\sum_{j=1}^{N}S_{j}^{x\left(y\right)}$.
The magnetic order parameter is $m=(m_{x}^{2}+m_{y}^{2})^{1/2}$,
and we define an associated correlation length $\xi^{\perp}$. To
discriminate between the states $\psi_{2}$ and $\psi_{3}$ we define
a clock-like order parameter $m_{6}=m\mbox{ cos}\left(6\varphi\right),$
where $\varphi=\mbox{tan}^{-1}\left(m_{y}/m_{x}\right)$, with $m_{6}$
being positive (negative) for the $\psi_{2\left(3\right)}$ states.
Moreover, we also keep track of the order-parameter distribution $P\left(m_{x},m_{y}\right)$~\cite{zhitomirsky14,lou07,wenzel11},
which is obtained considering statistics from different samples. In
order to study spin-glass freezing we use the spin-glass (Edwards-Anderson)
order parameter $q^{\alpha,\beta}\left(\mathbf{p}\right)=N^{-1}\sum_{j}S_{j}^{\alpha\left(1\right)}S_{j}^{\beta\left(2\right)}e^{i\mathbf{p}\cdot\mathbf{r}_{j}}$,
where $\alpha$ and $\beta$ are spin components, and $^{(1)}$ and
$^{(2)}$ denote two identical copies of the system (``replicas'')
containing the same defect configuration. We define the spin-glass
correlation length $\xi_{{\rm SG}}$ from this order parameter~\cite{pixley08},
and the freezing temperature $T_{f}$ is obtained locating the crossing
point of $\xi_{{\rm SG}}(T)/L$ for different system sizes $L$.
Note that the Edwards-Anderson parameter alone does not differentiate spin glass from LRO \cite{suppl}.

In the following we focus on the case of site dilution relevant, e.g.,
to the compound Er$_{2-x}$Y$_{x}$Ti$_{2}$O$_{7}$~\cite{gaudet16};
bond randomness is discussed in Ref.~\onlinecite{suppl}. Sample
results are presented in Fig.~\ref{fig:fig2}. We confirm the tendency
towards $\psi_{3}$ order at low $T$ and small $x$, see Figs.~\ref{fig:fig2}(b)
and (e) for $x=0.1$, in accordance with earlier simulations~\cite{maryasin14,andreanov15}.
We determine the transition from $\psi_{2}$ to $\psi_{3}$
as the point where the curve $m_{6}(T)$ changes sign for $L=12$,
see inset of Fig.~\ref{fig:fig2}(e). On general grounds, we
expect this transition to be first order. However, we find no traces
of it in the specific heat, as also reported in Refs. \cite{maryasin14, todoroki04},
a fact which contributes to the error bars in Fig.~\ref{fig:fig2}(a).
We leave a more detailed investigation of this point for future work.

For larger $x\gtrsim0.3$ the behavior becomes more glassy: This is seen in the specific-heat
curves, i.e., the maxima for all $L$ become broad and size-independent,
signaling the building up of short-range magnetic correlations above
a freezing temperature $T_{f}$~\cite{fischer}, Fig.~\ref{fig:fig2}(b).
Moreover, the correlation-length data $\xi^{\perp}(T)/L$
do not display crossing points, Fig. \ref{fig:fig2}(c), whereas $\xi_{{\rm SG}}(T)/L$
show well-defined crossing points, Fig. \ref{fig:fig2}(d). This signals
glassy freezing in the {\em absence} of LRO. The freezing temperature
$T_{f}(x)$ decreases with $x$, but remains finite up to the percolation
threshold, $x_{{\rm p}}=0.61$, Fig.~\ref{fig:fig2}(a). The fate
of the system is best judged by considering $m_{6}$ and the distribution
$P\left(m_{x},m_{y}\right)$, Figs.~\ref{fig:fig2}(e) and (f). $m_{6}$
is essentially zero for $x\gtrsim0.4$, and $P\left(m_{x},m_{y}\right)$
is peaked along a circle (instead of displaying sharp maxima), suggesting
coexisting finite domains with distinct local spin orientations.
We consider these MC data as clear evidence for the breaking
of the system into domains with frozen spin configurations and
no LRO, i.e., a CSG -- this is the central result of this letter.

The transition from $\psi_{2,3}$ to the CSG can be determined from the absence both of crossing points in $\xi^{\perp}(T)/L$ and of a clear-cut temperature trend of $m_6$. The resulting quantitative phase diagram is in Fig.~\ref{fig:fig2}(a), in remarkable agreement with the qualitative considerations which led to Fig.~\ref{fig:PD}(a).


\emph{Quantum effects.---} Turning to the quantum case, we note that the main effect of quantum
fluctuations is to stabilize $\psi_{2}$ even at $T=0$. This makes
$\psi_{2}$ more competitive against $\psi_{3}$ and shifts the corresponding
boundary to finite $x$ \cite{maryasin14}. As the competition between
$\psi_{3}$ and CSG is expected to be weakly affected by quantum fluctuations,
the extent of the $\psi_{3}$ phase consequently shrinks, Fig.~\ref{fig:PD}(b).
For strong quantum effects, $S=1/2$, we may speculate that $\psi_{3}$
disappears completely from the phase diagram, yielding a direct transition
from $\psi_{2}$ to CSG for all temperatures below $T_{f}$, but this
requires a more detailed and quantitative analysis of quantum effects which is beyond
the scope of this paper.


\emph{Experiments.---} We now confront our theory with experimental
data. Assuming that $m=\left\langle S^{x}\right\rangle =\frac{1}{2}$
and $\left\langle S^{y}\right\rangle =0$, we obtain the surprisingly
simple result $\delta h=\sqrt{3x\left(1-x\right)}J^{\pm\pm}$ for
site dilution~\cite{suppl}. For Er$_{2}$Ti$_{2}$O$_{7}$, extensive
investigations of the order-by-disorder mechanism estimate $J^{\pm\pm}=4.2\times10^{-2}\,{\rm meV}$
\cite{savary12} and $\Delta=5.3\times10^{-2}$\,meV~\cite{ross14}.
To discuss the disappearance of LRO, and given that Eq.~\eqref{crit}
is valid at weak disorder only, we instead compare the strength of
the fluctuations, $\delta h$, to the clean-limit spin gap, $\Delta$,
to obtain an upper bound for the critical randomness which reads $\delta \hc=f\Delta$
where $f$ is a numerical factor of order unity. Experimentally, LRO
disappears in Er$_{2-x}$Y$_{x}$Ti$_{2}$O$_{7}$ around $\xc\approx0.15$~\cite{gaudet16};
from which we extract $f\approx1/2$. Evidently, a more quantitative theory
is desirable for an accurate determination of $\xc$ -- this is left
for future work.
Our results are applicable
to Er$_{2}$Pt$_{2}$O$_{7}$ as well, a compound which shows so-called Palmer-Chalker
order $\left(\lam>2\right)$ \cite{hallas17}, not arising from an
order-by-disorder mechanism. Using $f=1/2$ and experimentally known model parameters, we predict that
a small amount of vacancies, $\xc\approx4\%$, destabilizes the magnetic
order in Er$_{2}$Pt$_{2}$O$_{7}$ \cite{suppl}.

\emph{Broader aspects of the theory.---} Our ideas are of generic relevance to magnets with strong spin-orbit
coupling. Key to our scenario is the presence of off-diagonal exchange
couplings in the microscopic Hamiltonian. These break spin-rotational
invariance, leading to an anisotropy gap in a clean ordered state. In the presence
of inhomogeneities, the off-diagonal couplings also produce random
fields \cite{fieldvsaniso} which compete with the gap (protecting the ordered state)
and favor CSG formation instead. The critical level of disorder where LRO disappears is, of course, material-specific.

For the 2D quantum-spin-liquid candidate YbMgGaO$_{4}$~\cite{li15,paddison17},
it has been argued that quenched disorder in the off-diagonal couplings is responsible for the 
destruction of LRO~\cite{li17,zhu17} due to a ``pinning-field'' mechanism~\cite{zhu17}. According to our scenario, 
this mechanism is akin to the random-field one which leads to CSG phase. In lower dimensions ($d<5/2$~\cite{parisi17}),
the CSG phase melts into a fluctuating cluster-paramagnet phase, in accordance with the observations in Ref.~\onlinecite{zhu17}.



\emph{Conclusions.---} Combining analytical arguments and large-scale
MC simulations, we have shown that defects induce fluctuating random
fields in XY pyrochlore antiferromagnets. These ultimately destroy
magnetic LRO, leading to a CSG phase beyond a critical
level of randomness. Our theory resolves the previous discrepancy
between theory and experiment, and is in semi-quantitative agreement
with experimental data on diluted Er$_{2}$Ti$_{2}$O$_{7}$.

We expect our ideas to motivate further studies into
the non-trivial role of randomness in magnets with strong spin-orbit coupling,
where the presence of off-diagonal exchange terms triggers a non-trivial
competition between anisotropy gap and random fields \cite{fieldvsaniso}.

\begin{acknowledgments}
We acknowledge instructive discussions with B. Gaulin, H.-H. Klauss,
K. Ross, R. Sarkar, and M. Zhitomirsky. ECA was supported by FAPESP
(Brazil) Grant No. 2013/00681-8 and CNPq (Brazil) Grant No. 302065/2016-4.
JAH was supported by CNPq Grant No. 307548/2015-5 and FAPESP Grants
No. 2015/23849-7 and No. 2016/10826-1. SR and MV were supported by
DFG SFB 1143.
\end{acknowledgments}


%

\end{document}


\title{Supplementary information for:\\
 Cluster-glass phase in pyrochlore XY antiferromagnets with quenched
disorder}

\author{Eric C. Andrade}

\affiliation{Instituto de F\'{i}sica de São Carlos, Universidade de São Paulo,
C.P. 369, São Carlos, SP, 13560-970, Brazil}

\author{José A. Hoyos}

\affiliation{Instituto de F\'{i}sica de São Carlos, Universidade de São Paulo,
C.P. 369, São Carlos, SP, 13560-970, Brazil}

\author{Stephan Rachel}

\affiliation{Institut für Theoretische Physik, Technische Universität Dresden,
01062 Dresden, Germany}

\affiliation{School of Physics, University of Melbourne, Parkville, Victoria 3010,
Australia}

\author{Matthias Vojta}

\affiliation{Institut für Theoretische Physik, Technische Universität Dresden,
01062 Dresden, Germany}

\date{\today}

\maketitle

\section{Variance of the local exchange field}

If we assume long-range order (LRO) which is uniform in the local
frames of the Hamiltonian (1) in the main text, with $\left|\alpha\right|<2$,
we can write the spin average as $\left\langle \mathbf{S}\right\rangle =\left\langle S^{x}\right\rangle \hat{x}+\left\langle S^{y}\right\rangle \hat{y}=m\hat{n}_{\parallel}$.
The local exchange field is then given by 
\begin{equation}
\mathbf{h}_{j}=\sum_{k=1}^{z}(J_{jk}^{xx}\left\langle S^{x}\right\rangle +J_{jk}^{xy}\left\langle S^{y}\right\rangle )\hat{x}+(J_{jk}^{yy}\left\langle S^{y}\right\rangle +J_{jk}^{xy}\left\langle S^{x}\right\rangle )\hat{y},\label{eq:local-h-uni}
\end{equation}
with the sum running over all the $z=6$ nearest-neighbor (NN) sites.
In the presence of random off-diagonal\textcolor{black}{{} exchange
couplings the local exchange field $\mathbf{h}_{j}=h_{j}^{\parallel}\hat{n}_{\parallel}+h_{j}^{\perp}\hat{n}_{\perp}$
is not parallel to the mean magnetization, where we defined the transverse
direction as $\hat{n}_{\perp}=\left(-\left\langle S^{y}\right\rangle \hat{x}+\left\langle S^{x}\right\rangle \hat{y}\right)/m$.
The general expression for the local transverse component is 
\begin{align}
\frac{h_{j}^{\perp}}{J^{\pm\pm}}= & \frac{2}{m}\sum_{k=1}^{z}\epsilon_{jk}\sin\theta_{jk}\left(\left\langle S^{x}\right\rangle ^{2}-\left\langle S^{y}\right\rangle ^{2}\right)\nonumber \\
 & -\frac{4}{m}\sum_{k=1}^{z}\epsilon_{jk}\cos\theta_{jk}\left\langle S^{x}\right\rangle \left\langle S^{y}\right\rangle .\label{eq:local-h}
\end{align}
} We now make two assumptions: (i) the magnetization $m$ is uniform
and (ii) $m=\left\langle S^{x}\right\rangle =\frac{1}{2}$ and thus
$\left\langle S^{y}\right\rangle =0$. Assumption (i) is motivated
by the fact that we are interested in the stability of the uniform
long-range order, and (ii) is made for simplicity. Equation~\eqref{eq:local-h}
then becomes 
\begin{equation}
u_{j}=\frac{h_{j}^{\perp}}{J^{\pm\pm}}=\sum_{k=1}^{6}\epsilon_{jk}\sin\theta_{jk},\label{eq:u}
\end{equation}
as presented in the main text. Here $\theta_{1}=\theta_{4}=0$, $\theta_{2}=\theta_{5}=\frac{2\pi}{3}$
and $\theta_{3}=\theta_{6}=-\frac{2\pi}{3}$.

For site dilution we have $\epsilon_{jk}=0$ if the neighboring site
$k$ is occupied, and $-1$ if vacant. Thus, $\epsilon_{jk}=0$ with
probability $(1-x)$ and $-1$ with probability $x$. For bond randomness,
on the other hand, $\epsilon_{jk}=+W$ if the coupling with the neighboring
site $k$ is the stronger one, and $-W$ (with $0<W<1$), for the
weaker one. Let $x$ be the concentration of weaker bond, then $\epsilon_{jk}=+W$
with probability $(1-x)$ and $-W$ with probability $x$. Here, the
similarity between bond randomness and site dilution becomes apparent.

We now calculate the possible values of $u$ and their corresponding
probabilities considering a fixed number of neighboring impurities
(from one to five) uniformly distributed (we drop the site index $j$
in the following discussion). For a single impurity we obtain three
values: $u=\pm\omega,\,0$, with $\omega=\frac{\sqrt{3}}{2}$ for
vacancies and $\omega=\sqrt{3}W$ for bond defects, all of them with
equal probability, i.e., $P_{1}(0)=P_{1}(\pm\omega)=\frac{1}{3}$.
For two impurities we then get $u=\pm2\omega,\,\pm\omega,\,0$ with
probabilities $P_{2}(\pm2\omega)=\frac{1}{15}$, $P_{2}(\pm\omega)=\frac{4}{15}$
and $P_{2}(0)=\frac{1}{3}$, respectively. Repeating this exercises
for three impurities, we find that $P_{3}(\pm2\omega)=\frac{1}{10}$,
$P_{3}(\pm\omega)=\frac{1}{5}$, and $P_{3}(0)=\frac{2}{5}$. For
4 and 5 impurities, we obtain the same results as for 2 and 1 impurities,
respectively, i.e., $P_{n}(u)=P_{6-n}(u)$. Since $P_{n}\left(u\right)=P_{n}\left(-u\right)$,
then $\overline{u}=0$. The variance is 
\begin{align}
\overline{u^{2}} & =\sum_{n=1}^{5}\sum_{u}u^{2}P_{n}(u)\times\left(\begin{array}{c}
6\\
n
\end{array}\right)x^{n}\left(1-x\right)^{6-n}\nonumber \\
 & =4x\left(1-x\right)\omega^{2},\label{eq:var_hperp_def}
\end{align}
where $\left(\begin{array}{c}
6\\
n
\end{array}\right)x^{n}\left(1-x\right)^{6-n}$ takes into account all possible ways of placing $n$ impurities among
the 6 neighboring sites (or sharing bonds). We remark that we are
not excluding sites which do not belong to the infinite cluster. This
leads to tiny corrections to~\eqref{eq:var_hperp}. Therefore, for
site dilution, we get 
\begin{equation}
\delta h=\sqrt{\overline{u^{2}}}J^{\pm\pm}=\sqrt{3x\left(1-x\right)}J^{\pm\pm},\label{eq:var_hperp}
\end{equation}
as stated in the main text, and 
\begin{equation}
\delta h=2\sqrt{3x\left(1-x\right)}WJ^{\pm\pm}
\end{equation}
for bond randomness.

We note that the effective anisotropy energy $\lambda$, describing
the mean-field selection of $\psi_{3}$, scales with $x$ in the case
of dilution and with $W^{2}$ in the case of bond disorder~\cite{maryasin14},
such that both cases are expected to display a similar competition
between $\psi_{3}$ selection and random-field destruction of order,
as captured by the stability criterion in Eq.~(4) in the main text.
However, in our numerical simulations for bond disorder, as described
in Sec.~\ref{sec:morenum} below, we have been unable to find the
$\psi_{3}$ state for small randomness in the temperature range investigated
(in contrast to the case of site dilution). Instead, we observed only
the $\psi_{2}$ and CSG states even at the classical level.


\section{Monte-Carlo simulation details}

Our Monte-Carlo (MC) simulations are performed on clusters with $N=4L^{3}$
spins with periodic boundary conditions and $L$ varying from $6$
to $12$. We employ three distinct types of MC moves: (a) single-site
(restricted) Metropolis updates, (b) microcanonical steps~\cite{alonso96}
and (c) parallel tempering~\cite{partemp}. Typically, we perform
$5\times10^{5}$ MC sweeps for thermalization, followed by $5\times10^{5}$
sweeps to calculate thermal averages. In our implementation, after
$10$ microcanonical sweeps we perform a Metropolis sweep followed
by a parallel tempering update. For the restricted Metropolis step,
we use a temperature-dependent selection window to ensure an average
acceptance rate larger than $50\%$ at any given temperature. Moreover,
we select our temperature grid such that a parallel tempering move
has a success rate larger than $40\%$. On top of thermal averages,
we also perform average typically from over $1\,000$ defects configurations
for $L=6$ down to $300$ configurations for $L=12$.

In our MC simulations the spins become frozen at low temperature,
with non-vanishing expectation values, $\left\langle \mathbf{S}_{i}\right\rangle \neq0$.
If all spins point along the same local direction, we then have $\mathbf{M}=N^{-1}\sum_{i}\left\langle \mathbf{S}_{i}\right\rangle \neq0$
which signals LRO. On the other hand, if the spins are frozen in a
disordered configuration as in a spin glass we have $\mathbf{M}=0$.
In order to detect spin freezing, independently of LRO, we introduce
the Edwards-Anderson order parameter~\cite{fischer,pixley08} 
\begin{eqnarray}
\tilde{q}^{\alpha,\beta} & = & \overline{N^{-1}\sum_{i}\langle S_{i}^{\alpha}\rangle\langle S_{i}^{\beta}\rangle},\label{eq:q_sg_def1}
\end{eqnarray}
where $\alpha,\beta=x,y,z$ are the spin components and the overline
denotes average over disorder. Within the MC simulation, the thermal
averages are replaced by averages over MC time, and we get 
\begin{eqnarray}
\tilde{q}^{\alpha,\beta} & = & \overline{N^{-1}\sum_{i}\mathcal{M}^{-2}\sum_{t_{1,}t_{2}}S_{i}^{\alpha}\left(t_{1}\right)S_{i}^{\beta}\left(t_{2}\right)},\label{eq:q_sg_MC1}
\end{eqnarray}
where $\mathcal{M}$ is the number of Monte Carlo steps. In general,
$S_{i}^{\alpha}\left(t_{1}\right)$ and $S_{i}^{\beta}\left(t_{2}\right)$
have different configurations, and it is then convenient to independently
simulate two copies of the system with identical defect configuration
(two replicas) 
\begin{eqnarray}
\tilde{q}^{\alpha,\beta} & = & \overline{\mathcal{M}^{-1}\sum_{t}N^{-1}\sum_{i}S_{i}^{\alpha\left(1\right)}\left(t\right)S_{i}^{\beta\left(2\right)}\left(t\right)}\\
 & = & \overline{\left\langle N^{-1}\sum_{i}S_{i}^{\alpha\left(1\right)}S_{i}^{\beta\left(2\right)}\right\rangle }.
\end{eqnarray}
Finally, it is convenient to introduce a $\mathbf{q}$-dependent Edwards-Anderson
order parameter 
\begin{eqnarray}
q^{\alpha,\beta}\left(\mathbf{q}\right) & = & \frac{1}{N}\sum_{i}S_{i}^{\alpha\left(1\right)}S_{i}^{\beta\left(2\right)}e^{i\mathbf{q}\cdot\mathbf{r}_{i}},\label{eq:q_sg_MC4}
\end{eqnarray}
which is the expression presented in the main text. The spin-glass
susceptibility is then given by 
\begin{eqnarray}
\chi_{SG}\left(\mathbf{q}\right) & = & \overline{N\sum_{\alpha,\beta}\left\langle \left|q^{\alpha,\beta}\left(\mathbf{q}\right)\right|^{2}\right\rangle }.\label{eq:chi_sg_def_k}
\end{eqnarray}


\section{Additional numerical results\label{sec:morenum}}

\subsection{Bond disorder}

In order to simulate bond defects, we consider a binary distribution
of bonds strengths described by $\epsilon_{jk}$ in Eq.~(3) of the
main text taking values $+W$ or $-W$ with equal probability. This
choice is inspired by the experimental situation in NaCaCo$_{2}$F$_{7}$~\cite{ross16,sarkar17}
and NaSrCo$_{2}$F$_{7}$~\cite{ross17}.

\begin{figure}[tb]
\begin{centering}
\includegraphics[width=0.9\columnwidth]{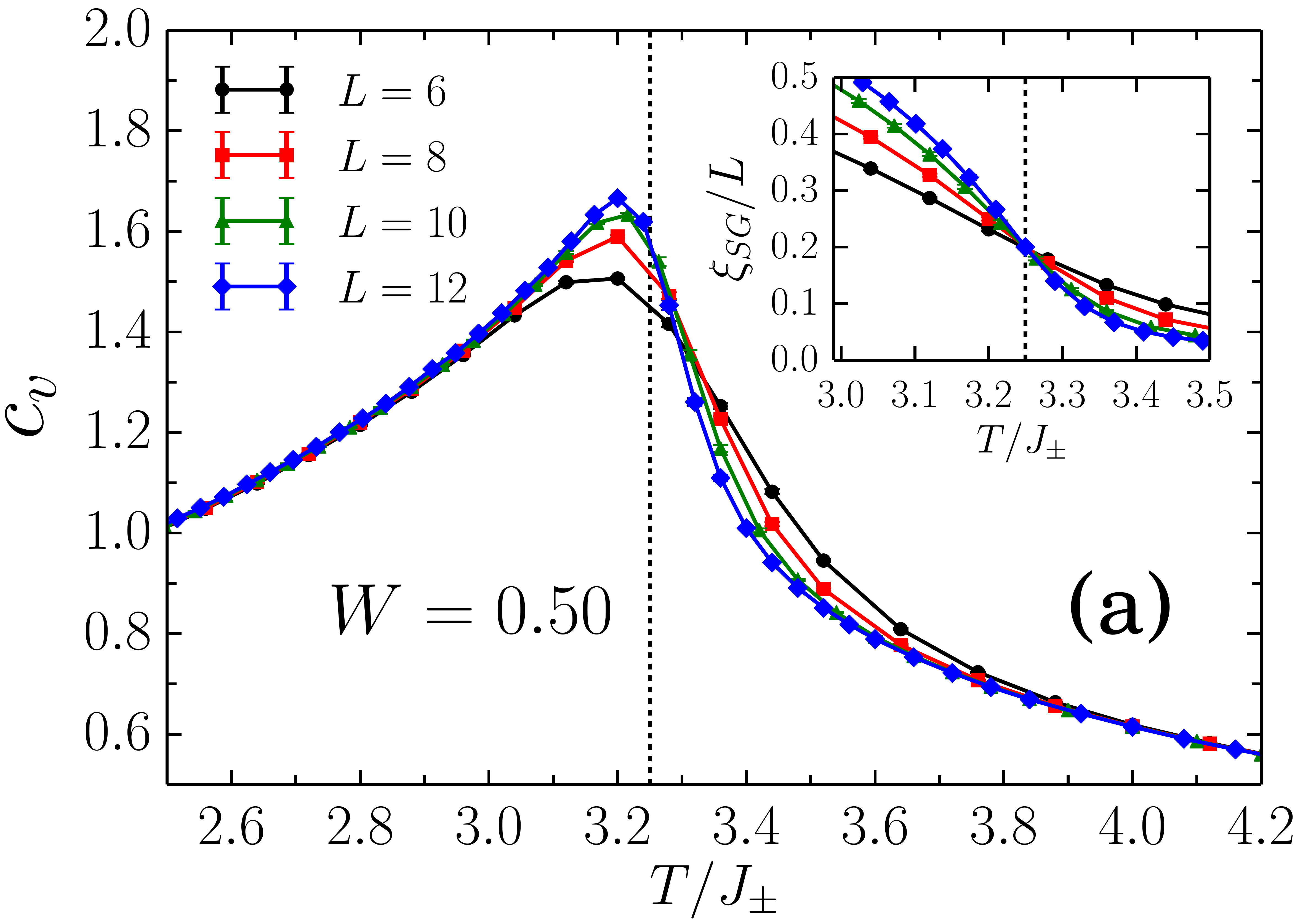} \\
\includegraphics[width=0.88\columnwidth]{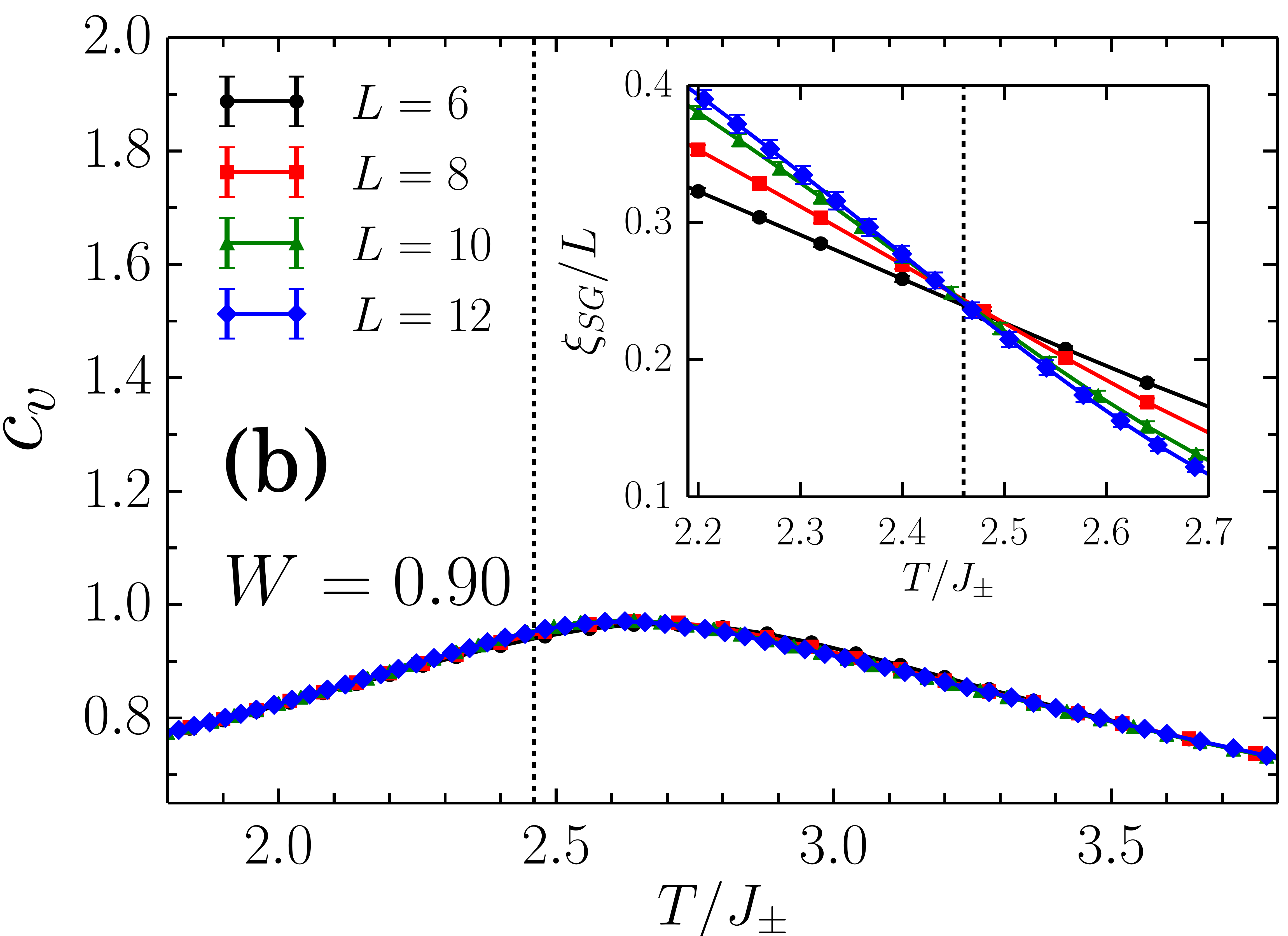} 
\par\end{centering}

\caption{Classical MC results for the Hamiltonian~(1) f the main text for
off-diagonal exchange ratio $\alpha=1$ and bond disorder. (a) Specific
heat $c_{v}$ as a function of temperature $T$ for bond disorder
$W=0.5$. (b) Same as (a), but for $W=0.9$. Inset: spin-glass correlation
length divided by the system size $\xi_{{\rm SG}}/L$ as a function
of $T$ showing a crossing point at $T_{f}/J_{\pm}=3.25(2)$ and $2.46(2)$
for $W=0.5$ and $0.9$, respectively. \label{fig:fig1s}}
\end{figure}

\begin{figure}[!t]
\begin{centering}
\hspace*{15pt}\includegraphics[width=0.9\columnwidth]{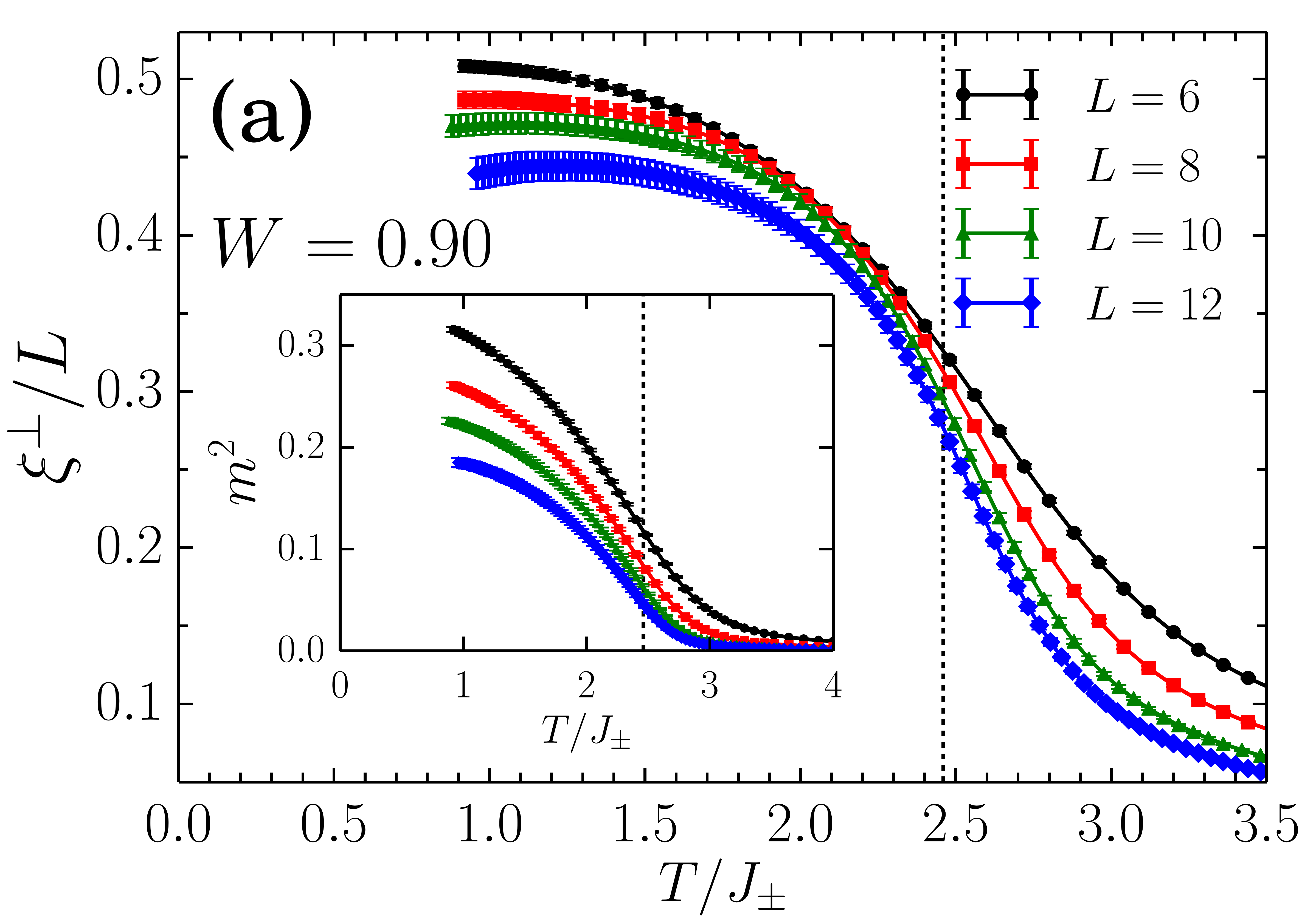}\\
\includegraphics[width=0.9\columnwidth]{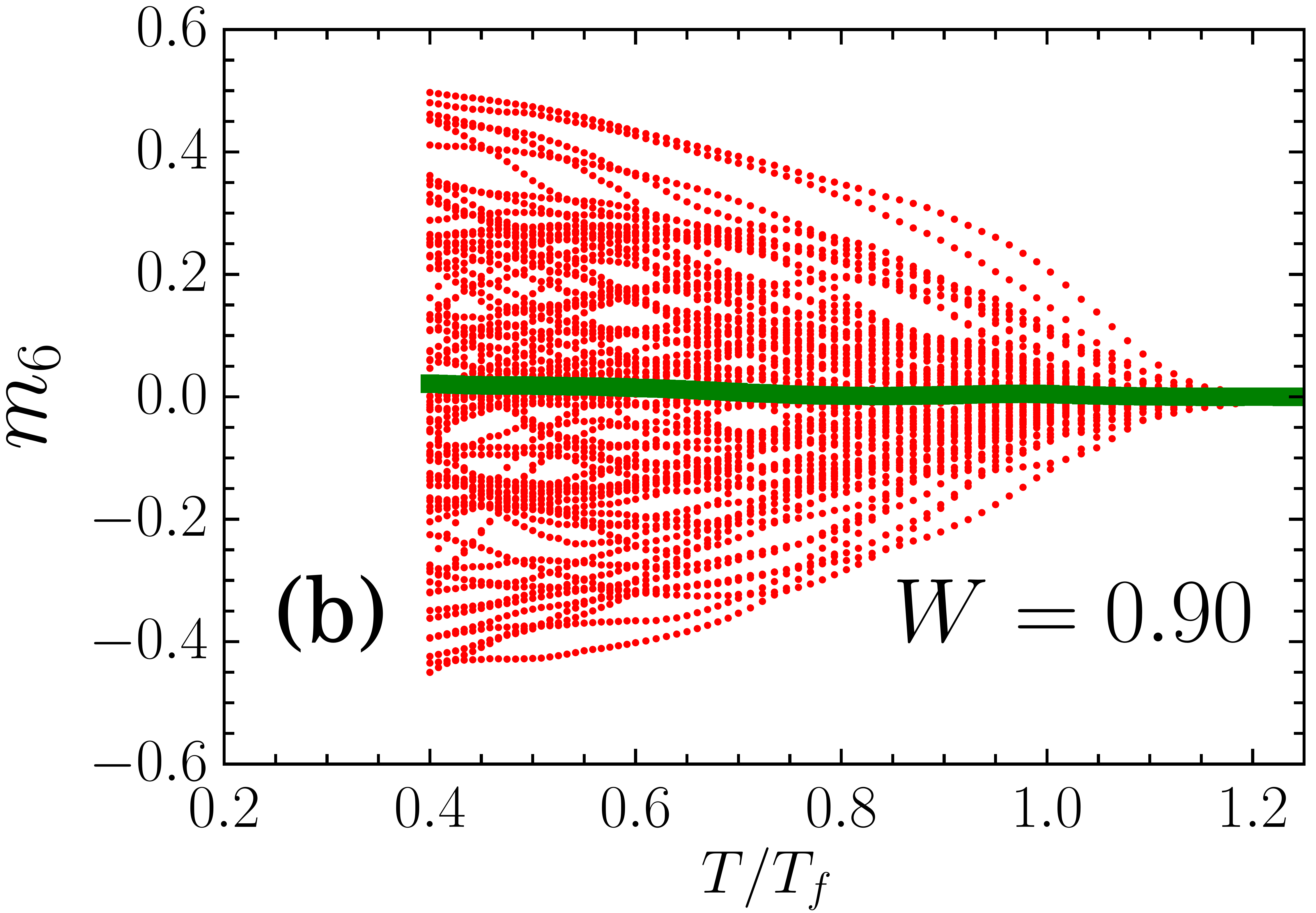}\\
\includegraphics[width=0.9\columnwidth]{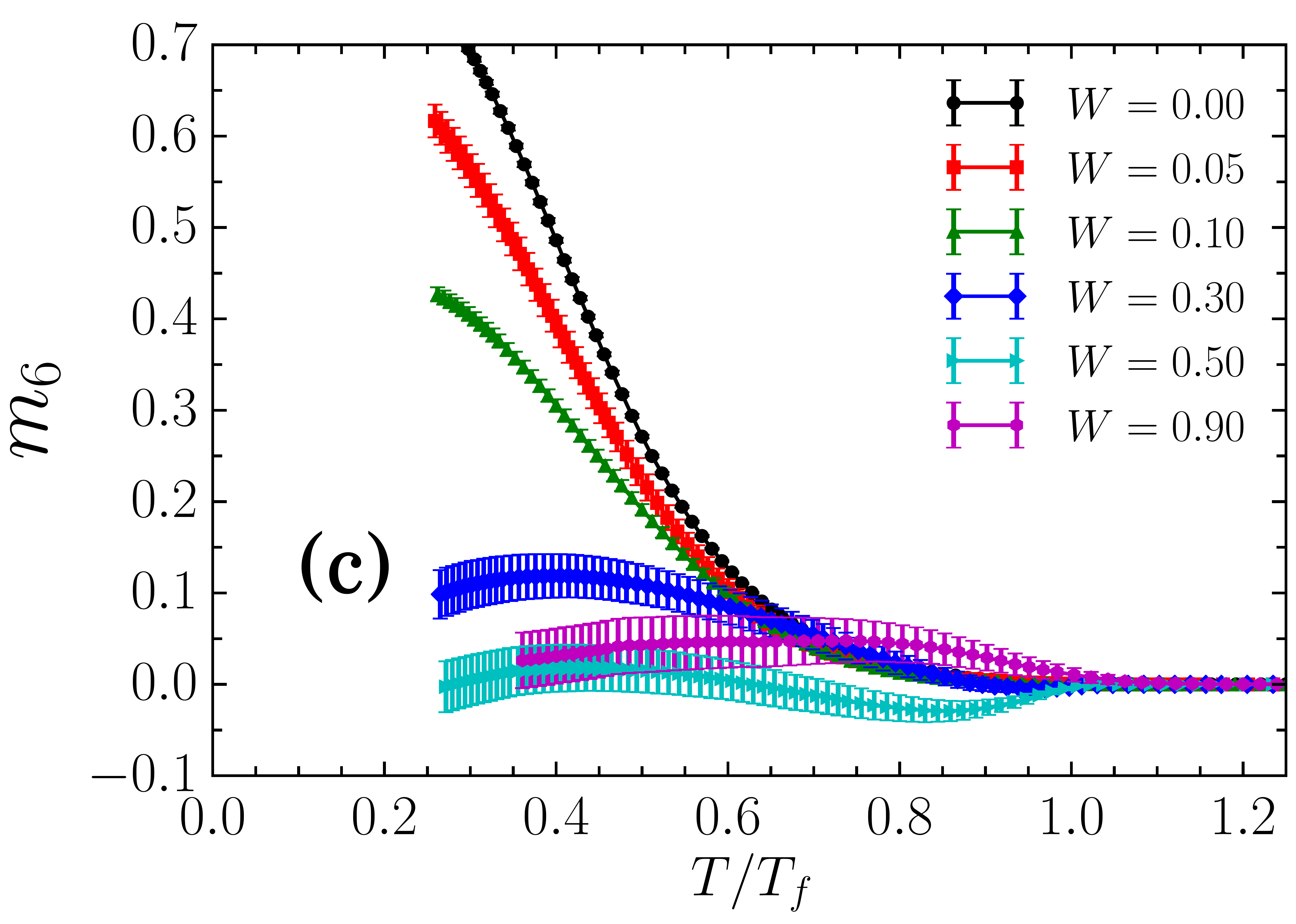} 
\par\end{centering}

\caption{Classical MC results for the Hamiltonian~(1) of the main text for
off-diagonal exchange ratio $\alpha=1$ and bond disorder. (a) Magnetic
correlation length plotted as $\xi^{\perp}(T)/L$ for $W=0.9$. Inset:
square of the magnetic order parameter as a function of $T$. (b)
Clock order parameter $m_{6}$ (green squares; error bars are smaller
than the symbol sizes) as a function of $T/T_{f}$ for $L=12$ and
$W=0.9$. The result for $80$ different defect configurations (red
dots) are also shown in order to illustrate the large sample-to-sample
fluctuations. (c) Clock order parameter $m_{6}$ as a function of
$T/T_{f}$ for five different bond disorder values and $L=10$. \label{fig:fig2s}}
\end{figure}

\begin{figure}
\begin{centering}
\includegraphics[width=0.85\columnwidth]{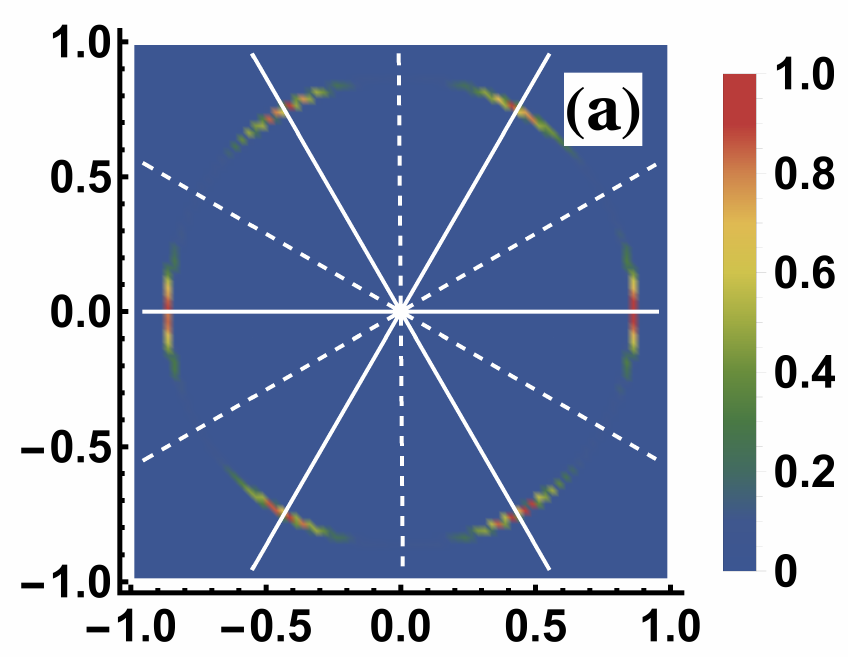}\\
\includegraphics[width=0.85\columnwidth]{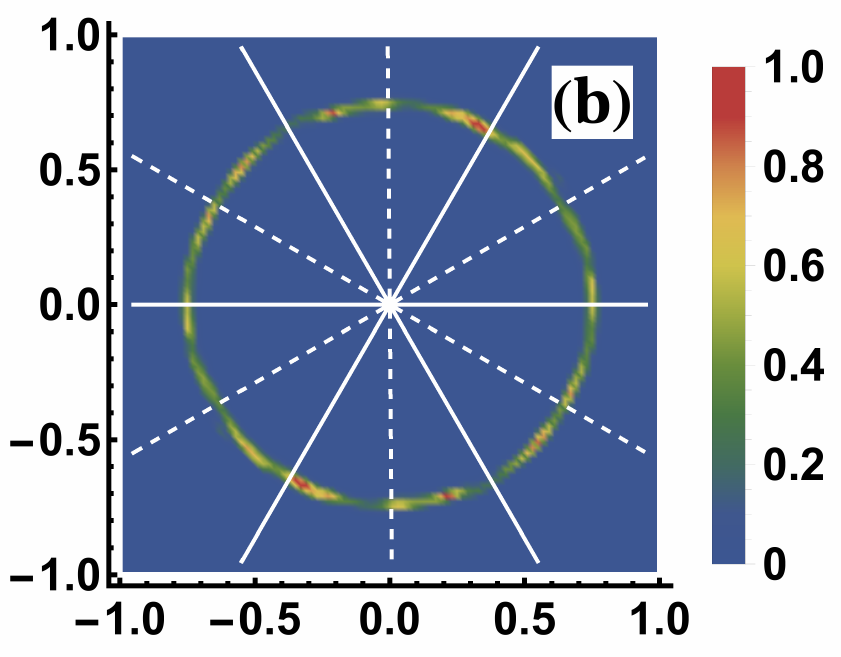}\\
\includegraphics[width=0.85\columnwidth]{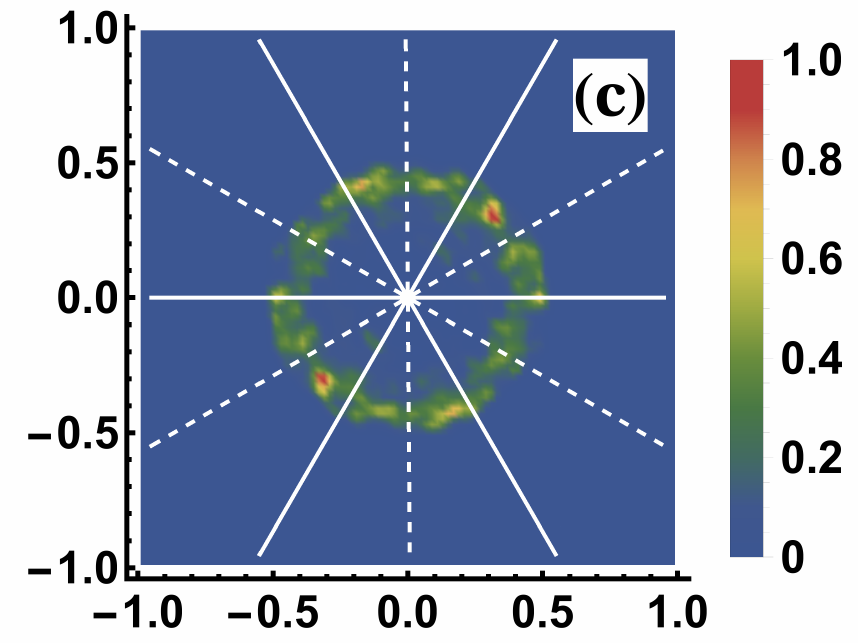}
\par\end{centering}

\caption{Sample-to-sample distribution function $P\left(m_{x},m_{y}\right)$
at $T=T_{f}/2$ for $L=12$ obtained via classical Monte Carlo simulation
of the Hamiltonian~(1) of the main text for off-diagonal exchange
parameter $\alpha=1$ with bond disorder (a) $W=0$ (the clean system,
for comparison), (b) $W=0.5$, and (c) $W=0.9$. The full (dashed)
radial lines show the expected positions of the peaks associated to
the $\psi_{2(3)}$ states. \label{fig:fig3s}}
\end{figure}

Sample results are displayed in Figs.~\ref{fig:fig1s}--\ref{fig:fig3s}.
The freezing temperature $T_{f}$ is obtained studying the crossing
point of the spin-glass correlation length $\xi_{SG}$ divided by
the system size $L$, as a function of $T$, as shown in the insets
of Figs.~\hyperref[fig:fig1s]{\ref{fig:fig1s}(a)} and \hyperref[fig:fig1s]{\ref{fig:fig1s}(b)}.
In Fig.~\ref{fig:fig1s} we clearly see that the specific heat becomes
more glass-like with increasing $W$, i.e., the maxima of $C(T)$
become broad and $L$-independent, characteristic of short-range magnetic
correlations building up above $T_{f}$~\cite{fischer}, as in the
case of site dilution. In Fig.~\hyperref[fig:fig2s]{\ref{fig:fig2s}(a)}
we present the magnetic correlation length $\xi^{\perp}$, divided
by the system size $L$, as a function of $T$ for strong bond disorder
$W=0.9$. The absence of a crossing point indicates the lack of magnetic
LRO, which is in accordance with the monotonic suppression of the
magnetic order parameter $m$ as $L$ increases {[}see the inset of
Fig.~\hyperref[fig:fig2s]{\ref{fig:fig2s}(a)}{]}. Moreover, $\xi^{\perp}/L$
saturates at low temperatures due to the formation of finite-size
domains. To further illustrate the glassy behavior, we show in Fig.~\hyperref[fig:fig2s]{\ref{fig:fig2s}(b)}
the clock-like order parameter $m_{6}=m\mbox{ cos}\left(6\varphi\right)$
as a function of $T$ for $80$ different bond defect configurations.
It is clear that no particular magnetically ordered state is selected
since $\overline{\left\langle m_{6}\right\rangle }\approx0$ in the
temperature window investigated (down to $T\approx T_{f}/2$). Therefore,
similar to the case of site dilution, a CSG phase emerges for large
structural defect concentration as the local random-field fluctuations
win over the LRO and break the system into domains.

For smaller disorder strength $W$ the situation appears, however,
somewhat different compared to the site-diluted case. Here, our Monte-Carlo
simulations find no evidence of a $\psi_{3}$ phase down to temperatures
$T_{f}/4$, see Fig.~\hyperref[fig:fig2s]{\ref{fig:fig2s}(c)}. For
$W\le0.10$ we always observe $\overline{\left\langle m_{6}\right\rangle }>0$,
with $\overline{\left\langle m_{6}\right\rangle }$ monotonically
decreasing as we increase $W$. This implies that the $\psi_{3}$
phase in the phase diagram, see Fig. 1(a) of the main text, either
exists only at very low temperatures or is absent entirely. In the
temperature range probed, we always observe a direct transition from
$\psi_{2}$ to CSG as $W$ increases. This apparent difference between
site dilution and bond disorder may be related to our specific choice
of binary bond disorder. Perhaps a different type of disorder distribution,
e.g. box disorder~\cite{maryasin14}, would be able to induce a more
prominent $\psi_{3}$ phase, although earlier MC simulations by some
of us also did not find it~\cite{sarkar17}. At this point we note
that, in principle, a different behavior of site and bond dilution
close to the homogeneous limit cannot be excluded~\cite{henley89},
since a missing site can be seen as $z=6$ NN weak bonds coupled to
a given site, a configuration which occurs with probability $\left(1/2\right)^{6}=0.016$
for the current choice of parameters. We postpone a more detailed
investigation of the weakly inhomogeneous regime, and the associated
peculiar response of a given defect distribution, to a future publication.

\begin{figure}[!t]
\begin{centering}
\includegraphics[width=0.88\columnwidth]{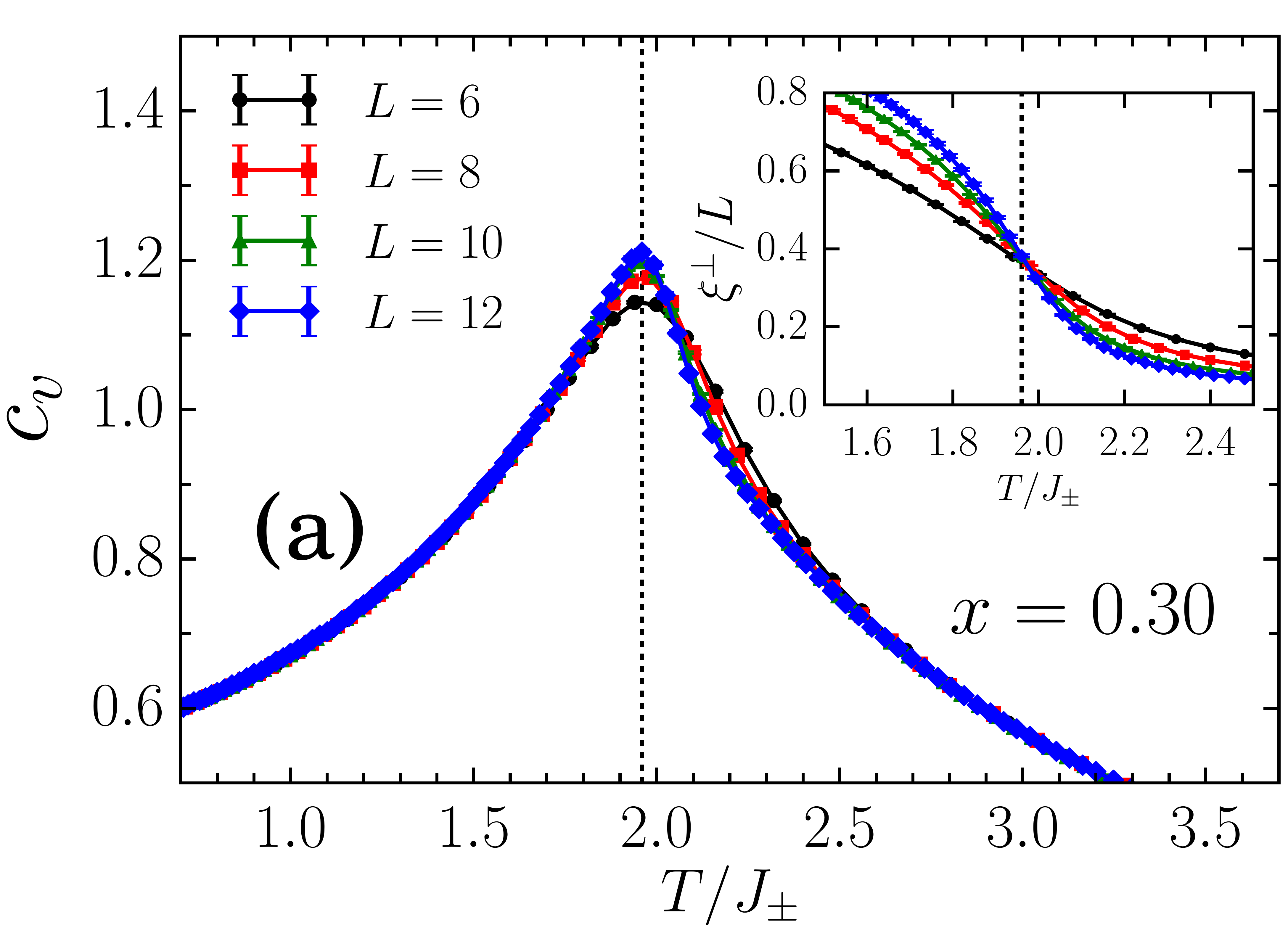}\\
\includegraphics[width=0.9\columnwidth]{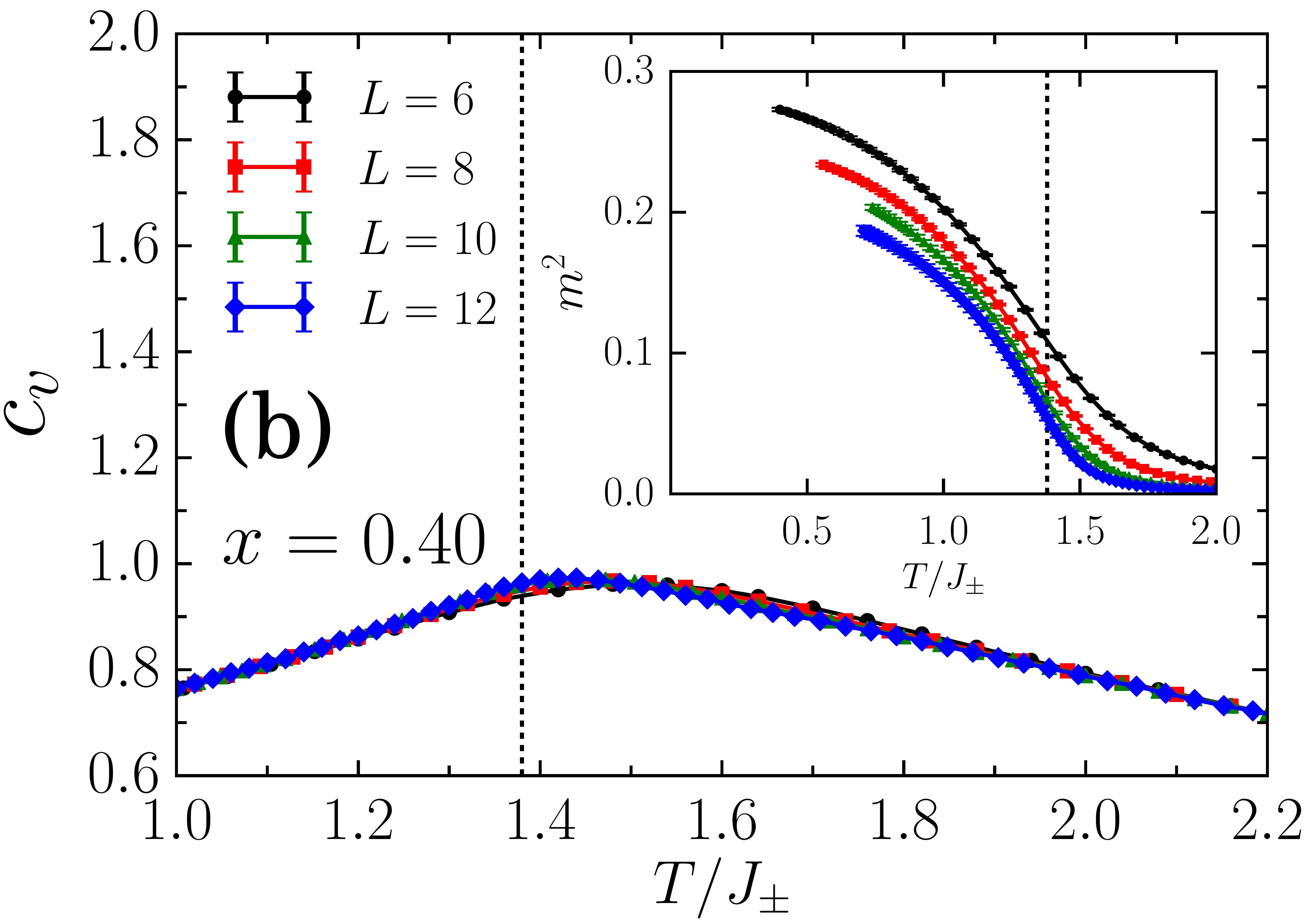}
\par\end{centering}

\caption{Classical MC results for the Hamiltonian~(1) for off-diagonal exchange
ratio $\alpha=1$ and various values of site dilution $x$. (a,b)
Specific heat $c_{v}$ as a function of temperature $T$ for $x=0.3$
and $x=0.4$, respectively. The vertical dashed line marks the position
of the freezing temperature $T_{f}$. Insets: (a) Magnetic correlation
length plotted as $\xi^{\perp}(T)/L$ for $x=0.30$, showing a crossing
point at $T_{f}/J_{\pm}=1.96(2)$. (b) Square of the magnetic order
parameter as a function of $T$ for $x=0.4$. \label{fig:fig4s}}
\end{figure}

The two-dimensional order-parameter distribution $P\left(m_{x},m_{y}\right)$
is shown in Fig.~\ref{fig:fig3s}. In the clean case, $W=0$, the
MC simulations nicely capture the six peaks corresponding to the $\psi_{2}$
state. In contrast, for large $W$ we observe the formation of clusters
corresponding to several magnetic states, not restricted to $\psi_{2}$
and $\psi_{3}$, and no well-separated sharp peaks are formed. These
findings are in accordance with our CSG scenario, since the (local)
magnetic order in the clusters is dictated by the local random anisotropies.

\subsection{Site dilution}

In Fig.~\ref{fig:fig4s} we show the specific heat curves as a function
of the temperature for the case of site dilution. For $x=0.3$ the
specific heat still displays a weak size dependence, and the magnetic
correlation length shows a crossing point, indicating that LRO survives,
albeit weakly. It eventually disappears around $x_{{\rm cr}}\approx0.35$.
As we increase the dilution to $x=0.4$ both the specific heat and
the magnetic order parameter curves exhibit the expected size dependence
of a CSG, Fig.~\ref{fig:fig4s}, which is in accordance with the
lack of crossing point in the correlation-length data $\xi^{\perp}(T)/L$
(Fig. 2(c) of the main text).

\begin{figure}[tb]
\begin{centering}
\includegraphics[width=0.85\columnwidth]{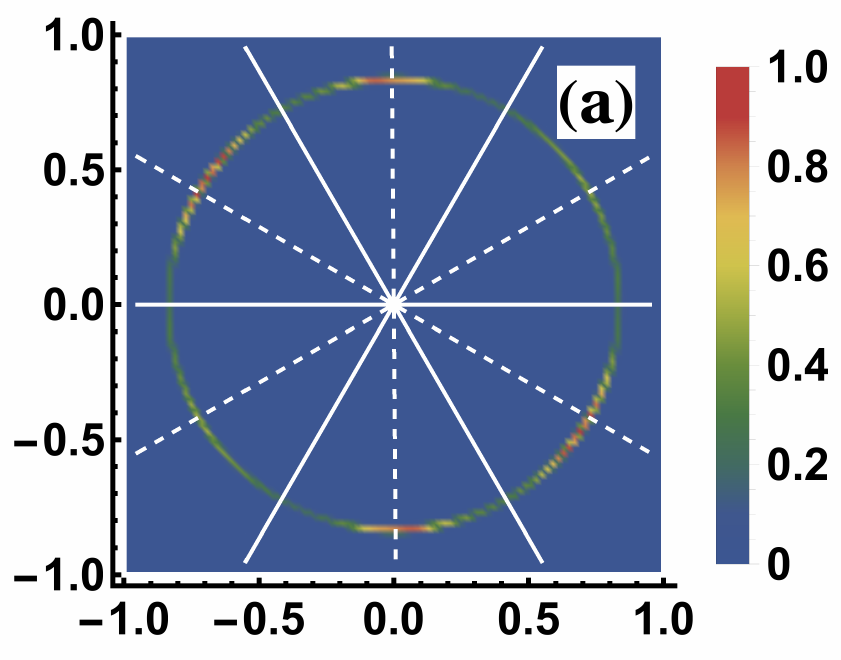}\\
\includegraphics[width=0.85\columnwidth]{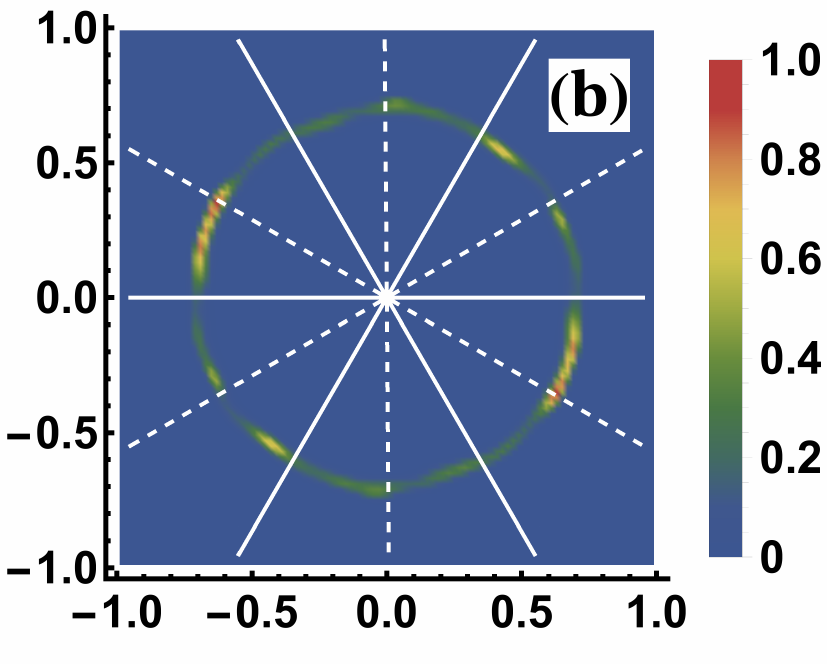}\\
\includegraphics[width=0.85\columnwidth]{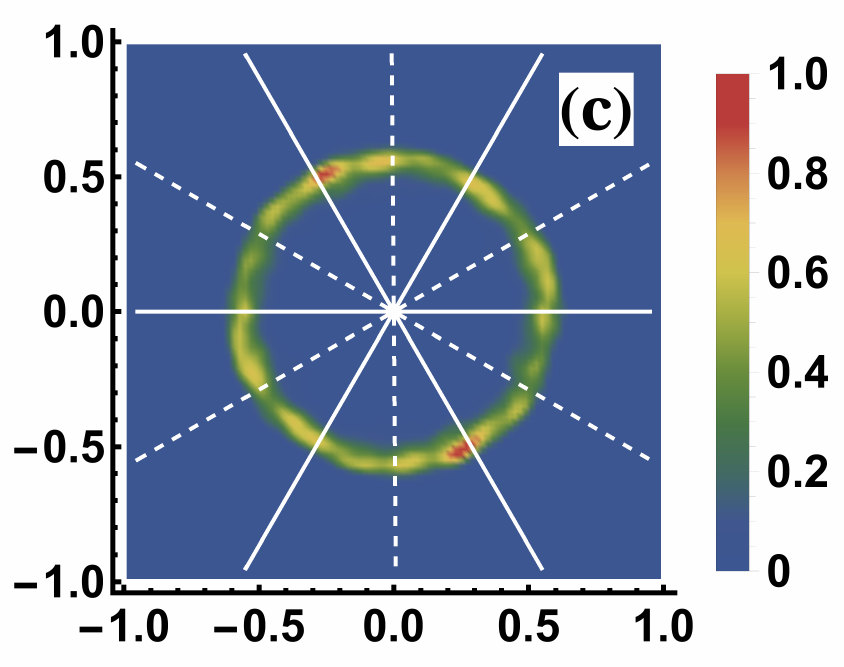}
\par\end{centering}

\caption{Same as Fig.~\ref{fig:fig3s} but for site dilution with concentrations
(a) $x=0.10$, (b) $x=0.30$, and (c) $x=0.40$. \label{fig:fig5s}}
\end{figure}

In Fig.~\ref{fig:fig5s} we present further results for the two-dimensional
order parameter distribution $P\left(m_{x},m_{y}\right)$, again in
the case of site dilution. For the smallest value of dilution studied,
$x=0.1$, our MC results are consistent with the selection of the
$\psi_{3}$ state, as reported in Refs.~\onlinecite{zhitomirsky14,andreanov15}
with $P\left(m_{x},m_{y}\right)$ displaying well-separated peaks
at the positions corresponding to $\psi_{3}$. In contrast, for larger
dilution, $x\gtrsim0.3$, we observe the formation of magnetic clusters
containing distinct magnetic states, with the (local) magnetic order
in theses clusters dictated by the local random anisotropies. This
is reflected in $P\left(m_{x},m_{y}\right)$ as it both shrinks, signaling
a reduction of $m$, and becomes peaked along a circle.

Summarizing, our classical MC simulations show, for both bond defects
and site dilution, the emergence of a CSG phase at large randomness.
The glass formation can be attributed to local random exchange fields,
which break the system into domains based on energetics.


\section{Stability criterion and comparison to experiments}

Here we give some details concerning the analytical calculations and
the stability estimates based upon them. We start with the effective
field theory employed by Savary \emph{et al}.~\cite{savary12} to
describe fluctuations within the classically degenerate manifold of
states. This theory is formulated in terms of an angle $\phi$ where
$\phi=\pi n/3$ ($\phi=\pi n/3+\pi/6$) corresponds to $\psi_{2}$
($\psi_{3}$) ordering, respectively. The action reads 
\begin{equation}
\mathcal{S}=\int\frac{d^{3}r}{v_{{\rm u.c.}}}d\tau\left[\sum_{\mu}\frac{\kappa_{\mu}}{2}(\partial_{\mu}\phi)^{2}+\frac{\eta}{2}(\partial_{\tau}\phi)^{2}-\frac{\tilde{\lambda}}{2}\cos6\phi\right].
\end{equation}
Here, $\kappa_{\mu}$ and $\eta$ are energies characterizing the
underlying gradient expansion, and $\tilde{\lambda}$ is an anisotropy
energy, for details see the supplement of Ref.~\onlinecite{savary12}.
Assuming $\tilde{\lambda}>0$ and expanding in fluctuations around
the minimum of the cosine, we find 
\begin{equation}
\chi^{\perp}(\mathbf{q},\omega)=\frac{1}{-\eta\omega^{2}+\kappa_{\mu}q_{\mu}^{2}+18\tilde{\lambda}}
\end{equation}
as the transverse susceptibility of the ordered $\psi_{2}$ state.
This matches the expression given in the main text, provided we take
the static limit and assign $\lambda=18\tilde{\lambda}$.

We now calculate the local transverse response due to fluctuating
transverse field. As outlined in the main text, we have 
\begin{equation}
\overline{\left\langle S_{i}^{\perp2}\right\rangle }=(\delta h)^{2}\int\frac{d^{3}q}{(2\pi)^{3}}\chi^{\perp}(\mathbf{q})^{2}.\label{intfluct}
\end{equation}
Power counting shows that the r.h.s. of this equation scales as $(\delta h)^{2}\kappa^{-3/2}\lambda^{-1/2}$
where $\kappa^{2}=\sum_{\mu}\kappa_{\mu}^{2}/3$ is the averaged gradient
energy. This eventually yields the stability criterion for weak disorder
advertised in the main text, Eq.~(5), here for $d=3$.

Since the LRO phase is stable for small disorder, any transition from
LRO to CSG has to appear at a finite level of disorder and is hence
beyond the linear-response theory used to derive~\eqref{intfluct}.
To estimate the critical level of disorder, we instead revert to a
qualitative criterion: Stability of LRO requires that the fluctuating
transverse field gap is smaller than the gap protecting the ordered
state. By using the gap of the homogeneous reference system, $\Delta=\sqrt{18\lambda/\eta}$,
we then obtain an \emph{upper bound} for the stability of LRO 
\begin{equation}
\delta h_{{\rm cr}}=f\Delta.\label{mf_hcr}
\end{equation}
Here, $\delta h_{{\rm cr}}$ is the critical level of randomness,
and simply $f$ is a numerical factor of order unity which we fix
by comparing to experimental results and/or numerical calculations.

We may benchmark the idea~\eqref{mf_hcr} using the 3d Ising model
in a random longitudinal field. Even though in this case there are
no transverse fluctuations, it is instructive to see if our arguments
hold here. The Imry-Ma criterion says that LRO is stable and therefore
$\delta h_{{\rm cr}}>0$. For a random field following a Gaussian
distribution, careful numerical calculations provide $\delta h_{{\rm cr}}=2.28\left(1\right)J$
at $T=0$~\cite{hartmann01}. If we use $\Delta=12J$ we obtain $f\approx1/5$
from Eq.~\eqref{mf_hcr}, suggesting that the criterion~\eqref{mf_hcr}
is reasonable.

We now apply this criterion to available experimental data; the results
of this comparison are given in the main text. For Er$_{2}$Ti$_{2}$O$_{7}$,
an extensive analysis of the order-by-disorder physics in the clean
limit has appeared in Ref.~\onlinecite{savary12}, yielding the estimate
$J^{\pm\pm}=4.2\times10^{-2}\,{\rm meV}$. Inserted into Eq.~\eqref{eq:var_hperp_def}
we obtain $\delta h$ as function of dilution level. This needs to
be compared to the clean-limit spin gap whose measured value is $\Delta=5.3\times10^{-2}$\,meV~\cite{ross14}.
Experimentally, LRO disappears in Er$_{2-x}$Y$_{x}$Ti$_{2}$O$_{7}$
around $x_{{\rm cr}}\approx0.15$~\cite{gaudet16}. Inserting this
into the criterion~\eqref{mf_hcr} and solving for $f$ we obtain
$f\approx1/2$, again suggesting that the criterion is reasonable.
With $f$ fixed for a class of systems, we can now use~\eqref{mf_hcr}
for other materials. 

In particular, our results should also be applicable to Er$_{2}$Pt$_{2}$O$_{7}$,
a compound which shows so-called Palmer-Chalker order as $\alpha>2$
\cite{hallas17}, not arising from an order-by-disorder mechanism.
It is easy to see that our simple expression for $\delta h$ also
holds in this case. From the phase diagram summarized in Ref.~\onlinecite{yan17}
and the transition temperatures for Er$_{2}$Ti$_{2}$O$_{7}$ and
Er$_{2}$Pt$_{2}$O$_{7}$ ($1.2\,{\rm K}$ and $0.38\,{\rm K}$,
respectively), we extract $\alpha\approx2.5$ and $J^{\pm\pm}\approx2.1\times10^{-1}\,{\rm meV}$
for Er$_{2}$Pt$_{2}$O$_{7}$. Using this estimative together with
the experimentally reported value of the spin gap, $\Delta=1.4\times10^{-1}\,{\rm meV}$~\cite{hallas17},
and again $f=1/2$, we predict that a small amount of vacancies, $x_{{\rm cr}}\approx4\%$,
destabilizes the magnetic order in diluted Er$_{2}$Pt$_{2}$O$_{7}$.
We note that, compared to Er$_{2}$Ti$_{2}$O$_{7}$, the larger value
of the anisotropic exchange ratio $\alpha=J^{\pm\pm}/J^{\pm}$ enhances
the effects of randomness.